\documentclass[twocolumn]{aastex631}
\usepackage[mathscr]{euscript}
\usepackage{multirow}

\def\muG{{\mu\rm G}}

\received{}
\revised{}
\accepted{}

\submitjournal{The Astrophysical Journal}

\shorttitle{Particle Acceleration at FR-II Jets}
\shortauthors{Seo et al.}

\begin{document}

\title{A Simulation Study of Ultra-relativistic Jets - III. Particle Acceleration at FR-II Jets}
\author[0000-0002-5550-8667]{Jeongbhin Seo}
\affiliation{Department of Physics, College of Natural Sciences, UNIST, Ulsan 44919, Korea}
\affiliation{Department of Earth Sciences, Pusan National University, Busan 46241, Korea}
\author[0000-0002-5455-2957]{Dongsu Ryu}
\affiliation{Department of Physics, College of Natural Sciences, UNIST, Ulsan 44919, Korea}
\author[0000-0002-4674-5687]{Hyesung Kang}
\affiliation{Department of Earth Sciences, Pusan National University, Busan 46241, Korea}
\correspondingauthor{Dongsu Ryu}\email{dsryu@unist.ac.kr}

\begin{abstract}


We study the acceleration of ultra-high-energy cosmic rays (UHECRs) at FR-II radio galaxies by performing Monte Carlo simulations for the transport, scattering, and energy change of the CR particles injected into the time-evolving jet flows that are realized through relativistic hydrodynamic (RHD) simulations. Toward that end, we adopt physically motivated models for the magnetic field and particle scattering. By identifying the primary acceleration process among diffusive shock acceleration (DSA), turbulent shear acceleration (TSA), and relativistic shear acceleration (RSA), we find that CRs of $E\lesssim1$ EeV gain energy mainly through DSA in the jet-spine flow and the backflow containing many shocks and turbulence. After they attain $E\gtrsim$ a few EeV, CRs are energized mostly via RSA at the jet-backflow interface, reaching energies well above $10^{20}$ eV. TSA makes a relatively minor contribution. The time-asymptotic energy spectrum of escaping particles is primarily governed by the jet power, shifting to higher energies at more powerful jets. The UHECR spectrum fits well to the double-power-law form, whose break energy, $E_{\rm break}$, corresponds to the size-limited maximum energy. It is close to $d\mathcal{N}/dE\propto E^{-0.5}$ below $E_{\rm break}$, while it follows $d\mathcal{N}/dE\propto E^{-2.6}$ above $E_{\rm break}$, decreasing more gradually than the exponential. The power-law slope of the high-energy end is determined by the energy boosts via non-gradual shear acceleration across the jet-backflow interface and the confinement by the elongated cocoon. We conclude that giant radio galaxies could be major contributors to the observed UHECRs.

\end{abstract}

\keywords{acceleration of particles --- cosmic rays --- galaxies: jets --- methods: numerical --- relativistic processes}

\section{Introduction}\label{s1}

The origin of ultra-high-energy cosmic rays (UHECRs) with energy $E\gtrsim$ EeV ($=10^{18}$ eV) remains by and large unknown. The potential accelerators are likely to be of extragalactic origin, because UHECRs have a Larmor radius too large to be confined magnetically within our Galaxy \citep[see][for reviews]{torres2004,batista2019}. Relativistic jets from active galactic nuclei (AGNs), characterized by a bulk Lorentz factor of {$\Gamma_j\sim 1-10$}, a magnetic field of $B\sim 100 \muG$, and a length scale of $L\sim 100$ kpc, are considered highly feasible sources of UHECRs \citep[see][for reviews]{blandford2019,rieger2019,hardcastle2020,matthews2020,eichmann2022}.

Acceleration scenarios of UHECRs, relevant to AGN jets, have been extensively investigated in previous studies, including the first-order Fermi (Fermi-I) acceleration (diffusive shock acceleration) mainly at sub-relativistic shocks in the jet-induced backflow \citep[e.g.,][]{matthews2019}, the stochastic second-order Fermi (Fermi-II) acceleration by turbulent flows in the lobe \citep[e.g.,][]{hardcastle2010}, the gradual shear acceleration in relativistic shearing flows \citep[e.g.,][]{rieger2004,webb2018,rieger2019}, the discrete shear acceleration at the interface between the jet-spine and backflow \citep[e.g.,][]{ostrowski1998,kimura2018}, the turbulent shear acceleration \citep{ohira2013}, and the espresso mechanism \citep{caprioli2015,mbarek2019,mbarek2021}. 

Using a newly developed relativistic hydrodynamic (RHD) code \citep[][hereafter Paper I]{seo2021a}, we recently performed RHD simulations of FR-II type jets, which propagate up to several tens of kpc into an intracluster medium (ICM) of ``constant density'' \citep[][hereafter Paper II]{seo2021b}. Models with broad ranges of jet parameters were considered: the jet Lorentz factor, $\Gamma_j\approx 2-70$, the jet power, $Q_j\approx 3\times10^{45}-3\times10^{47}{\rm erg~s^{-1}}$, the jet-to-background density contrast, $\eta\equiv\rho_j/\rho_b\approx 10^{-5}-10^{-3}$, and the jet-to-background pressure contrast, $P_j/P_b\approx1-10$. Here, $\Gamma_j=(1-(u_j/c)^2)^{-1/2}$ is specified by the inflow velocity of the jet, $u_j$, and $c$ is the speed of light. As shown in many previous simulation studies \citep[e.g.,][]{hardcastle2013,english2016,li2018,perucho2019}, we found that the overall jet morphology is governed primarily by $Q_j$; more powerful jets tend to develop narrower, more elongated lobes (cocoons), whereas less powerful jets have broader lobes full of shocks and turbulence. The interfaces between the jet-spine flow and backflow and also between the backflow and shocked ICM become turbulent via the Kelvin-Helmholtz instability due to strong velocity shear.

In Paper II, we also quantified nonlinear flow structures, such as shocks, turbulence, and velocity shear, generated in the jet-induced flows. Shocks in the jet-spine flow have relativistic speeds, $\beta_s=u_s/c \sim 0.2 - 1$, and sonic Mach numbers, $M_s \lesssim 5$, whereas those in the backflow are mildly relativistic with $\beta_s \sim 0.01 - 0.4$ and have $M_s \lesssim 2$. The relativistic shear coefficient, $\mathcal{S}_{r} = \Gamma_z^4\Omega_{\rm shear}/15$, which is inversely proportional to the timescale of relativistic shear acceleration \citep{webb2018}, is large, extending up to $\sim10^3-10^5(c/r_j)^2$, in the jet-spine flow, while its probability distribution function (PDF) peaks at $10^{-3}-10^{-2}(c/r_j)^2$ in the backflow. Here, $\Omega_{\rm shear}$, $\Gamma_z$, and $r_j$ are the velocity shear, the Lorentz factor of shear flows, and the jet radius, respectively. The turbulence generated in the jet-spine flow and backflow follows the Kolmogorov spectrum of $\propto k^{5/3}$ for $k\gtrsim2\pi/r_j$. The jet kinetic energy is dissipated through shocks, turbulence, and shear in the jet-spine flow and backflow, implying that the processes involving those nonlinear dynamics could be important in the production of UHECRs.

As a sequel to Paper II, in this paper, we investigate the acceleration of UHECRs in FR-II type jets. Specifically, through Monte Carlo (MC) simulations, we follow the transport, scattering, and energy change of the cosmic ray (CR) particles injected into the flows of simulated jets. For it, we perform RHD simulations to generate realistic FR-II type jets in the ``stratified ICM'', propagating them up to a few hundred kpc, and save a series of evolving snapshots of the jet-induced flows. CRs with initial energies of $E\sim 10^{13-15}$ eV are injected into the simulation domain, and their trajectories and scatterings are followed in a random walk fashion, as a ``post-processing'' step. Particles are assumed to interact with the magnetohydrodynamic (MHD) fluctuations frozen into underlying turbulent plasma flows, which are described based on a physically motivated model. They scatter according to a model recipe with a prescribed mean-free-path (MFP), $\lambda_f(E)$. 

A net energy change, $\Delta E$, arises as a result of the Lorentz transformation between the moving fluid frame and the simulation (laboratory) frame. In the MC simulations, particles encounter numerous shocks, chaotic turbulent flows, and shear flows. Thus, the acceleration processes can be categorized into the three main types: diffusive shock acceleration (DSA), turbulence shear acceleration (TSA), and relativistic shear acceleration (RSA). We attempt to estimate the relative importance of the different types. Finally, we quantify the resulting energy spectrum of escaping CRs in the time-asymptotic limit.

The MC technique was previously applied to investigate the acceleration of UHECRs in radio jets. \citet{ostrowski1998} and \citet{kimura2018}, for instance, considered simplified, static jet-cocoon systems with discontinuous shear of mildly relativistic flow; particles go through random walks via isotropic scattering, interacting with the underlying shear flow. While the details, such as particle injection and the prescription for diffusive motions of particles, are different, both found that CRs are efficiently energized to UHECRs via discrete shear acceleration, reaching $E\gtrsim10^{20}$ eV. The resulting energy spectrum of UHECRs is hard, whereas the cutoff at high energies is slower than the exponential\footnote{In Figure \ref{f8}(a), the power-law decrease of the energy spectrum at high energies is compared with the exponential cutoff.}. In addition, \citet{caprioli2015} proposed that CRs can gain energy via the so-called espresso scenario. Later, \citet{mbarek2019,mbarek2021} confirmed it in MHD jet simulations combined with particle orbit-propagation calculations.

Our numerical approach differs from these previous studies in a number of aspects including the followings: (1) A high-order accurate RHD code with a fully relativistic equation of state is employed to simulate the relativistic flows of radio jets, where shocks and turbulence, as well as shear, are well reproduced. (2) For the estimation of the magnetic field strength in the jet-induced flows, models based on known physics are considered. (3) In the random walk transport of CR particles, physically motivated, energy-dependent $\lambda_f$ and ``restricted'' scattering angles, $\delta\theta\leq\delta\theta_{\max}$ with energy-dependent $\delta\theta_{\max}$, are adopted. (4) The trajectories of CRs are integrated utilizing the time-evolving snapshot data taken from RHD jet simulations. (5) Considering the acceleration timescales of different processes, the contributions of DSA, TSA, and RSA to the energization of CR particles are evaluated. (6) Inspecting the CR trajectories, we examine the pathways through which the highest energy CRs gain energy, and then analyze how the high-energy end of the UHECR spectrum is produced.

The paper is organized as follows: In Section \ref{s2}, we describe the RHD simulations of FR-II type radio jets. In Section \ref{s3}, we discuss the basic physics of the particle acceleration processes that are expected to occur in jet-induced flows. In Section \ref{s4}, we describe MC simulations of particle transport, scattering, and acceleration. The results are presented in Section \ref{s5}. A brief summary is given in Section \ref{s6}.

\begin{deluxetable*}{cccccccccccc}[t]
\tablecaption{Simulation Models for FR-II Jets\label{t1}}
\tabletypesize{\small}
\tablecolumns{11}
\tablenum{1}
\tablewidth{0pt}
\tablehead{
\colhead{Model name$^a$} &
\colhead{$Q_j(\rm{erg~s^{-1}})$} &
\colhead{$\eta\equiv\frac{\rho_j}{\rho_b}$} &
\colhead{$\dot{M}_j(\rm{dyne})$} &
\colhead{$u_{j}/c$} &
\colhead{$\Gamma_{j}$} &
\colhead{$u_{\rm head}^*/c$} &
\colhead{$t_{\rm cross}(\rm{yrs})$} &
\colhead{Grid zones} &
\colhead{${N_j\equiv\frac{r_j}{\Delta x}}^b$} &
\colhead{${\frac{t_{\rm end}}{t_{\rm cross}}}$}
}
\startdata
Q45-$\eta5$-S &3.34E+45&1.E-05&1.15E+35&0.9905&7.2644&0.0409 &7.97E+4&$(400)^2\times600$ & 5 & 176\\
Q46-$\eta5$-H &3.34E+46&1.E-05&1.13E+36&0.9990&22.5645&0.1180 &2.77E+4&$(800)^2\times1200$ & 10 & 111\\
Q46-$\eta5$-S &3.34E+46&1.E-05&1.13E+36&0.9990&22.5645&0.1180 &2.77E+4&$(400)^2\times800$ & 5 & 200\\
{Q46-$\eta5$-L} &3.34E+46&1.E-05&1.13E+36&0.9990&22.5645&0.1180 &2.77E+4&$(240)^2\times360$ & 3 & 150\\
Q47-$\eta5$-S &3.34E+47&1.E-05&1.12E+37&0.9999&71.0149&0.2965 &1.10E+4&$(400)^2\times1000$ & 5 & 196\\
\hline
\enddata
\tablenotetext{^a}{The character ``S'' in the three models with different $Q_j$ stands for the density stratification in the background ICM; the stratification is given in Equation (\ref{kings}) with $\beta_K=0.5$, $r_c=50$ kpc, and $\rho_c=2.34\times10^{-27}\rm{g~cm^{-3}}$. {Those attached with ``H'' and ``L'' are the high and low-resolution models, respectively.}} 
\tablenotetext{^b}{Simulation resolution: the number of grid zones in the jet radius of $r_j=1$ kpc.}
\end{deluxetable*}

\section{RHD Simulations of Radio Jets}\label{s2}

Recently, we developed a three-dimensional (3D) RHD code based on the 5th-order accurate, finite-difference WENO (weighted essentially non-oscillatory) scheme \citep{jiang1996,borges2008} and the 4th-order accurate, SSPRK (strong stability preserving Runge–Kutta) time discretization \citep{spiteri2003}. The RC version of the equation of state \citep{ryu2006} was incorporated in order to correctly reproduce the thermodynamics across the relativistic fluid in the jet and the nonrelativistic ICM. The details of other numerical implementations and tests for the code can be found in Paper I. In Paper II, we simulated the dynamical evolution of AGN jets of FR-II type, ejected into a ``uniform'' medium of the ICM density, using this newly developed RHD code.

For the current study, we perform simulations for jets into a more realistic, ``stratified'' ICM background, propagating them to larger distances, as listed in Table \ref{t1}.

\subsection{Setup for Stratified ICM}\label{s2.1}

{Since we are interested in FR-II radio galaxies ejected into the ICM of galaxy clusters, we employ the so-called King profile to emulate the density stratification in the typical cluster core region:}
\begin{equation}
\rho_{\rm ICM}(r) = \rho_{c}\left[ 1+ \left( \frac{r}{r_{c}}\right)^{2}\right]^{-3\beta_K/2},
\label{kings}
\end{equation}
where $r$ is the distance from the cluster center, and $r_c$, $\rho_c$, and $\beta_K$ are the core radius, the core density, and a model parameter. respectively. We adopt $\beta_K=0.5$ and $r_c=50$ kpc. The hydrogen number density at the cluster center is set to be $n_{\rm H,ICM} = 10^{-3}$ cm$^{-3}$ as in Paper II, so $\rho_c=2.34\times10^{-27}\rm{g~cm^{-3}}$. Furthermore, the cluster core is assumed to be isothermal with $T_{\rm ICM} = 5\times10^7$ K. {In such a numerical set-up, we concentrate on the relatively late evolution of the jet as it propagates far away from the much denser environment near the central AGN.} To balance the pressure gradient due to the stratified background, an external gravity is imposed, and hydrostatic equilibrium is achieved without a jet.

The length and time scales of jet simulations are normalized with $r_0=r_j$ and $t_0=r_j/c$, respectively. Below, the simulation results are expressed in units of the following normalization\footnote{In this paper, the variables $u$, $P$, and $\mathcal{E}= \Gamma^{2}\rho h -P$ are used to represent the fluid velocity, pressure, and energy density, respectively, while $v$, $p$ and $E$ represent the particle velocity, momentum, and energy, respectively.}: $\rho_0=\rho_c$, $u_0=c$, and $P_0=\rho_0 c^2=2.1\times10^{-6}~{\rm erg~cm^{-3}}$. Then, the pressure at the cluster core corresponds to $P_c/P_0=7.64\times10^{-6}$ in dimensionless units.

The simulation domain is represented by a rectangular box in the 3D Cartesian coordinate system. The cluster center is located at the middle of the bottom surface, defined as $(x,y,z)=(0,0,0)$. A circular jet nozzle with a radius of $r_j$=1 kpc, through which the jet material is injected along the $z$-direction, is positioned at the cluster center. The outflow boundary condition is imposed on the bottom surface except for the jet nozzle. The continuous boundary condition is applied to the other five sides of the simulation domain.

\subsection{Setup for Jet Inflow}\label{s2.2}

The jet model can be specified by the jet power, $Q_j$, the density ratio, $\eta\equiv\rho_j/\rho_c$, and the pressure ratio, $P_j/P_c$, in our simulations. We consider only the models with $\eta=10^{-5}$ and $P_j=P_c$, and examine the acceleration of UHECRs in representative jets with different $Q_j$. The jet power is the rate of the energy inflow through the jet nozzle, excluding the mass energy:
\begin{equation}
Q_j=\pi r^{2}_ju_{j}\left(\Gamma^{2}_j\rho_jh_j-\Gamma_j\rho_jc^2\right),
\label{Qjet}
\end{equation}
where $h_j= (e_j + P_j)/\rho_j$ and $e_j$ are the specific enthalpy and the sum of the internal and rest-mass energy densities of the jet inflow, respectively. With fixed $r_j$, $\rho_j$, and $P_j$, $Q_j$ is determined by $u_j$ or $\Gamma_j$.

Table \ref{t1} lists the models, showing the ranges of jet-flow quantities: $Q_j\approx 3.34\times (10^{45}-10^{47})~{\rm erg~s^{-1}}$, $u_j/c\approx0.9905-0.9999$, and $\Gamma_j \approx7.3-71$. They intend to include the characteristic values inferred for observed FR-II radio jets \citep{ghisellini2001, godfrey2013}. The first column shows the model name, following the nomenclature of Paper II. The first two elements are the exponents of $Q_j$ and $\eta$. For instance, Q45-$\eta5$-S is for the model with $Q_j\approx 3.34\times 10^{45}~{\rm erg~s^{-1}}$ and $\eta=10^{-5}$. The character ``S'' appended in three models emphasizes the density ``stratification'' in the background ICM, while ``H'' and ``L'' denote ``high'' and ``Low'' resolutions. The grid resolution is given by the number of grid zones in the jet radius of $r_j=1$ kpc, $N_j = r_j/\Delta x$, in the 10th column. Here, $\Delta x$ is the size of grid zones. The 4th column shows the momentum injection rate or the jet thrust in Equation (10) of Paper II.

\begin{figure*}[t] 
\centering
\vskip 0.1 cm
\includegraphics[width=0.95\linewidth]{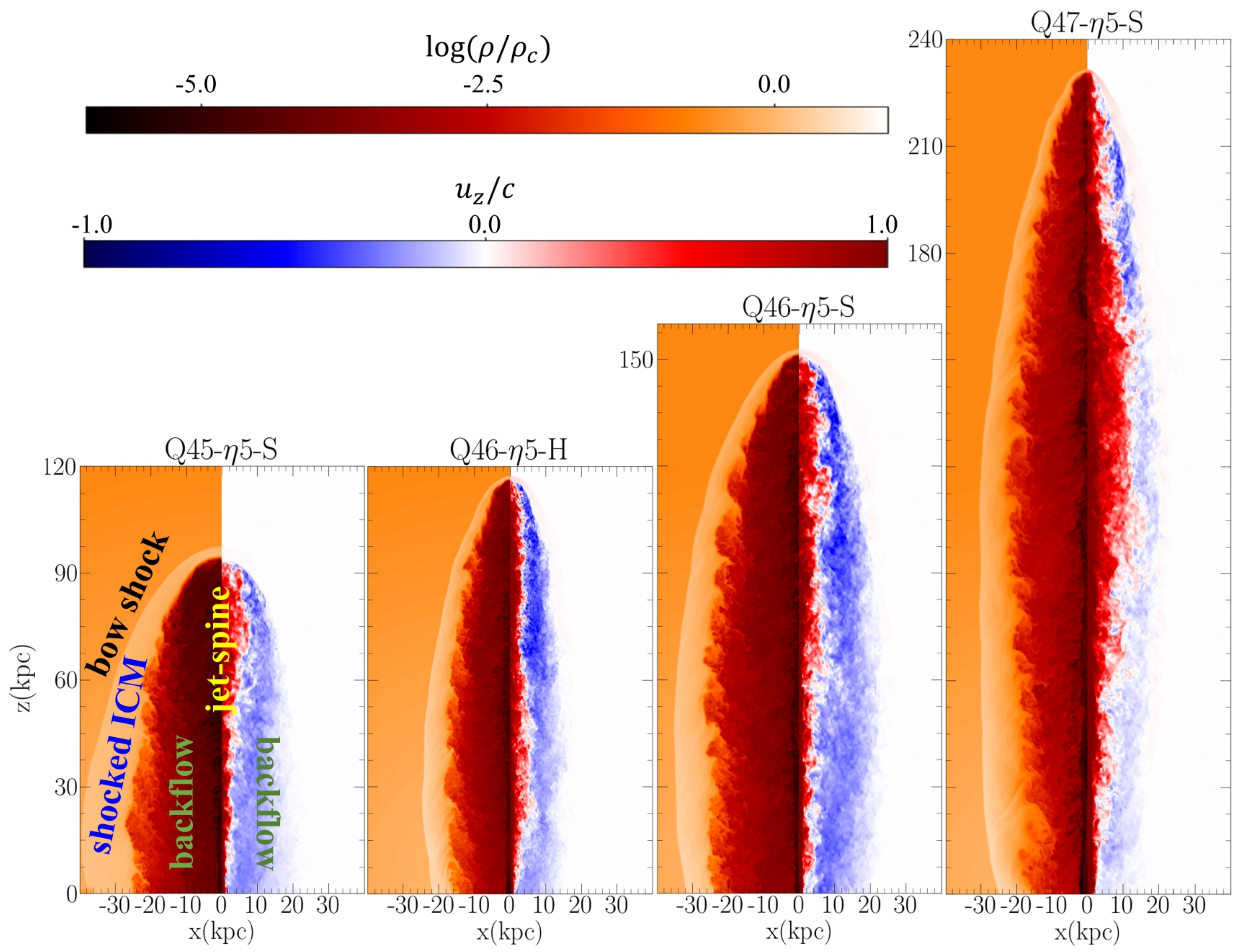}
\vskip -0.1 cm
\caption{2D slice images of the density, $\log\rho$, (left) and the $z$-velocity along the jet direction, $u_z$, (right) for the jet models considered. The model parameters are given in Table \ref{t1}, and the images are at $t=t_{\rm{end}}$. The important features of the jet-induced flows are marked in the leftmost panel. Note that while the jet-spine flow has positive vertical velocities (red), the backflow of the cocoon has negative vertical velocities (blue).}
\label{f1}
\end{figure*}

The dynamics of relativistic jets are commonly described with the jet head speed, $u_{\rm head}^*=u_j\cdot \sqrt{\eta_{r}}/(\sqrt{\eta_{r}}+1)$ \citep{marti1997}, which represents the approximate advance speed of the jet head derived from the balance between the jet ram pressure and the background pressure, and the jet crossing time given as $t_{\rm cross} = {r_j}/{u_{\rm head}^*}$. Here, $\eta_{r} = (\rho_{j}h_{j}\Gamma^{2}_j)/(\rho_c h_c)$ is the relativistic density contrast. The 7th and 8th columns of Table \ref{t1} show $u_{\rm head}^*$ and $t_{\rm cross}$, respectively. The end time of simulations, $t_{\rm end}/t_{\rm cross}$, is given in the last column. We note that simulations run up to $t_{\rm end}/t_{\rm cross}\sim110-200$, longer than those in Paper II, so the bow shock reaches up to a few hundred kpc (see Figure \ref{f1}), intending to reproduce realistic jet flows for the study of UHECR acceleration.

As in Paper II, a slow, small-angle precession with period $\tau_{\rm prec}=10~t_{\rm{cross}}$ and angle $\theta_{\rm prec}=0.5^{\circ}$ is added to the jet inflow velocity in order to break the rotational symmetry of the system.

\subsection{Simulated Jets}\label{s2.3}

The typical morphology of relativistic jets that is realized by numerical simulations may include the following features: recollimation shocks formed in the jet-spine flow, a terminal shock at the head of the jet, a cocoon of the shocked jet material flowing backward, the shocked ICM, and a bow shock that encompasses the entire jet-induced flow, \citep[e.g.,][and Paper II]{english2016, li2018,perucho2019}. Figure \ref{f1} shows the two-dimensional (2D) slice images of the density and the $z$-velocity along the jet direction for all the models considered, illustrating some of these features. Similar images for the models in the uniform ICM were presented in Figures 2 and 3 of Paper II. As can be seen in these figures, the jet morphology depends on the jet power $Q_j $; higher $Q_j$'s induce more elongated jets, while lower $Q_j$'s result in more extended, turbulent cocoons.

\begin{figure*}[t]
\centering
\vskip 0.1 cm
\includegraphics[width=0.95\linewidth]{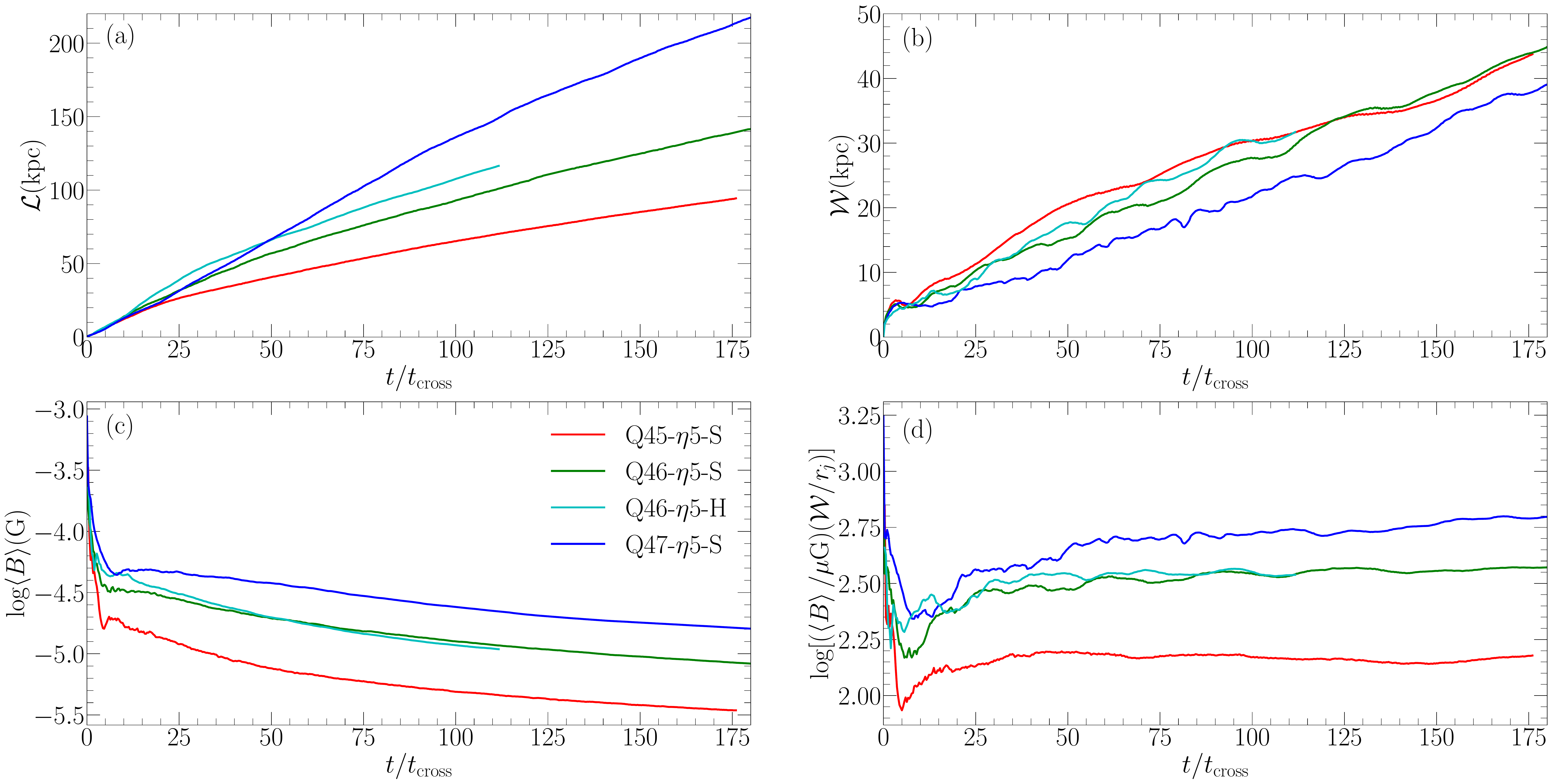}
\vskip -0.1 cm
\caption{Time evolution of the following quantities of the ``cocoon'' as a function of $t/t_{\rm cross}$ in simulated jets: (a) the length, $\mathcal{L}$, (b) the lateral width at its midpoint, $\mathcal{W}$, (c) the volume-averaged value of the comoving magnetic field strength in Equation (\ref{bmodel}), $\langle B \rangle$, (d) $\langle B \rangle\mathcal{W}$, which is used to estimate the size-limited $E_{\rm max}$ in Equation (\ref{Emax}).}
\label{f2}
\end{figure*}

The dynamics of jets in the stratified background models are overall similar to those in the uniform background models. While the latter are described in detail in Paper II, we here briefly comment the effects of density stratification, by examining the evolution of the length $\mathcal{L}$ and width $\mathcal{W}$ of the cocoon as a function of $t/t_{\rm cross}$, shown in panels (a) and (b) of Figure \ref{f2}; $\mathcal{L}/\mathcal{W}$ and $\mathcal{W}$ for the models in the uniform ICM are shown in Figure 5 of Paper II. We find that in the low-power model of Q45-$\eta5$-S, the jet advance is a little slower in the stratified background model, owing to the enhanced lateral expansion near the jet head, than in the uniform background model. By contrast, for the high-power models of Q46-$\eta5$-S and Q47-$\eta5$-S, the jet head penetrates a little faster into the density-decreasing ICM in the stratified background models. On the other hand, as the cocoon expands, although its width fluctuates differently in different models, $\mathcal{W}$ is on average almost identical in the models with stratified and uniform backgrounds.

We also find that, with the higher numerical resolution in Q46-$\eta5$-H, $\mathcal{L}$ and $\mathcal{W}$ are a bit larger than in Q46-$\eta5$-S. In addition, as we also showed in Paper II, nonlinear structures such as shocks, turbulence, and velocity shear are better captured in higher resolution simulations. As a consequence, the Monte Carlo simulations of CR acceleration would be somewhat resolution-dependent; the acceleration would be more efficient in Q46-$\eta5$-H than in Q46-$\eta5$-S (see Section \ref{s5.2} for the discussion on the resolution dependence of particle acceleration).

\begin{figure*}[t]
\centering
\vskip 0.1 cm
\includegraphics[width=0.95\linewidth]{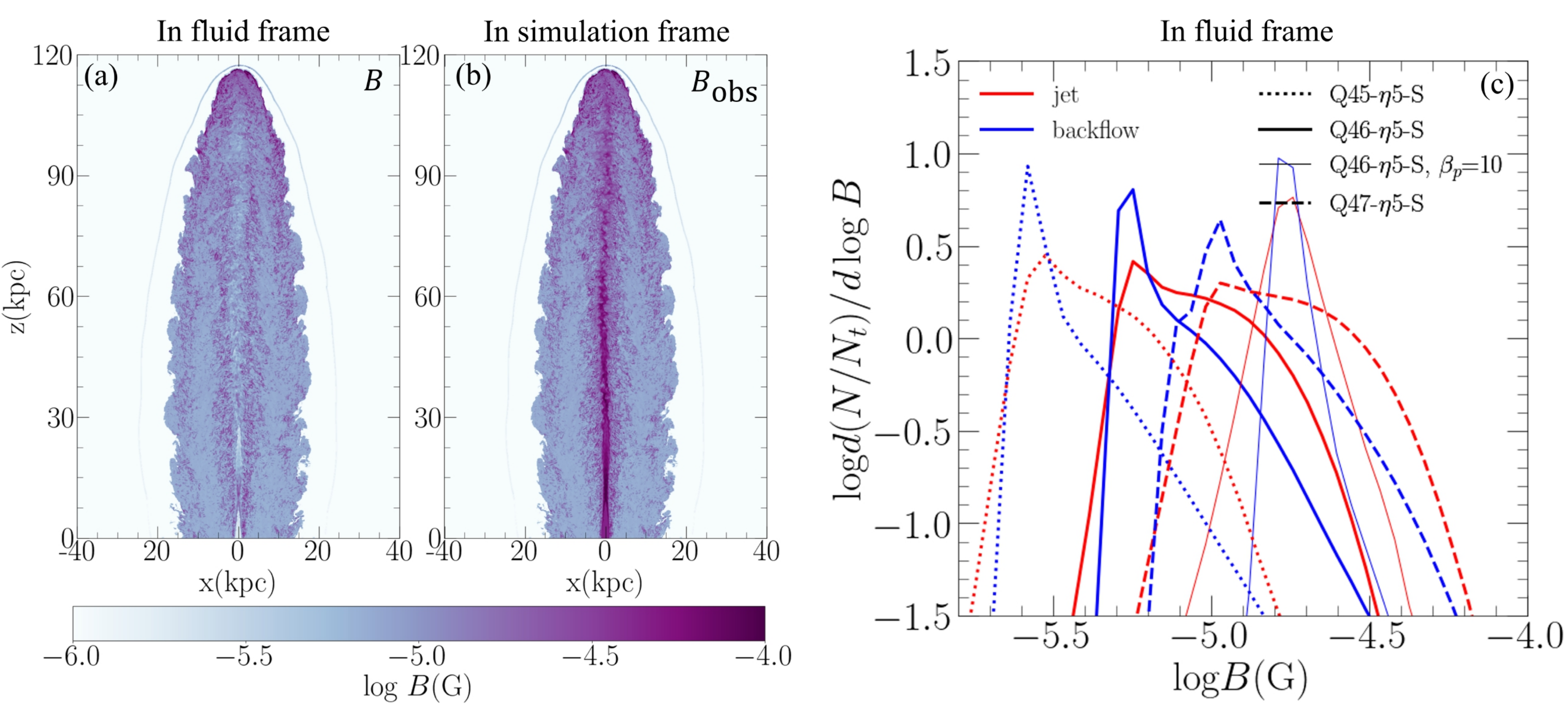}
\vskip -0.1 cm
\caption{Panels (a) and (b): 2D slice images of the magnetic field strength estimated with $\beta_p=100$ for the high-resolution model Q46-$\eta5$-H, the ``comoving'' magnetic field strength $B$ in the fluid frame (a) and the ``observing'' magnetic field strength $B_{\rm obs}\approx\Gamma_f B$ (with the fluid Lorentz factor $\Gamma_f$ ) in the simulation frame (b). Panel (c): {the probability distribution functions (PDFs)} of $B$ (estimated with $\beta_p=100$) in the jet-spine flow (red) and the backflow (blue) for the three ``S'' models (thick lines). The PDFs of $B$ estimated with $\beta_p=10$ in the Q46-$\eta5$-S model are also plotted for comparison (thin solid lines). All are shown at $t_{\rm end}$.}
\label{f3}
\end{figure*}

\begin{figure*}[t] 
\centering
\vskip 0.1 cm
\includegraphics[width=0.95\linewidth]{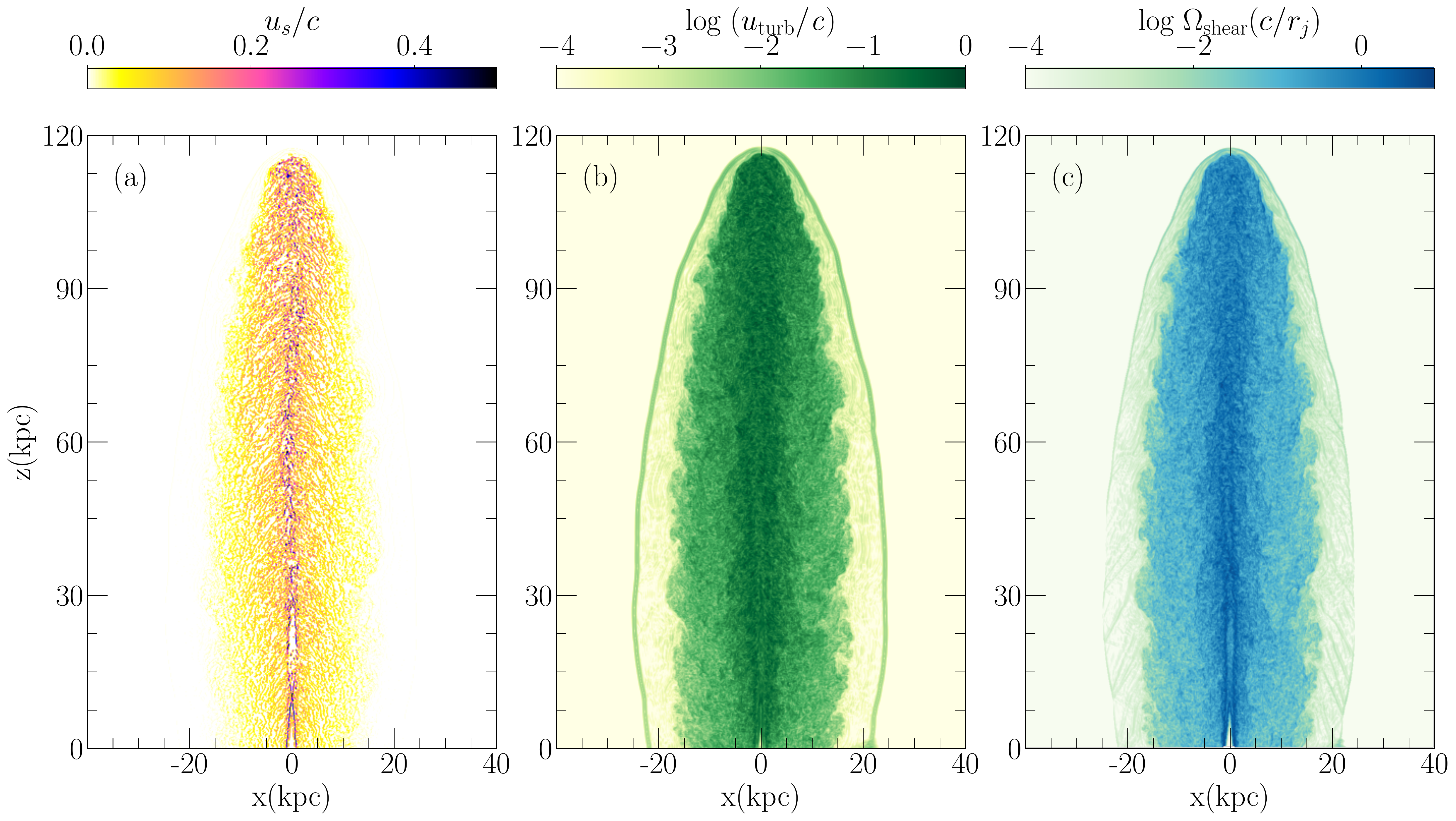}
\vskip -0.1 cm
\caption{2D slice images of the quantities that exhibit nonlinear flow dynamics: (a) shock speed, $u_s$, (b) turbulent flow velocity, $u_{\rm turb}$, and (c) velocity shear, $\Omega_{\rm shear}$. The jet of Q46-$\eta5$-H is shown at $t=t_{\rm{end}}$.}
\label{f4}
\end{figure*}

\subsection{Modeling of Magnetic Field}\label{s2.4}

While the magnetic field is one of the key physical ingredients that govern the particle acceleration processes, this paper is based on RHD simulations, partly because fully relativistic magnetohydrodynamic (RMHD) simulations are more challenging and require higher resolutions \citep[see, e.g.,][for a review]{marti2019}. Thus, we here adopt a set of prescriptions for the strength of magnetic field, $B$, which utilize the hydrodynamic properties of simulated jet flows, such as the internal energy, the turbulent energy, and the shock speed.

In the estimation of synchrotron emission in simulated radio jets, $B$ was often modeled assuming that the magnetic energy density is a fixed fraction of the internal energy density \citep[see, e.g.,][]{wilson1983,gomez1995}. Following the approach, we first parameterize $B$ with the plasma beta, $\beta_p=P/P_B$ as follows:
\begin{equation}
P_B=\frac{B_{p}^{2}}{8\pi} = \frac{P}{\beta_p}
\label{bie}.
\end{equation}
Here, $B_p$ denotes the magnetic field strength derived from the $\beta_p$ prescription. We adopt $\beta_p=100$ as the fiducial value and also consider $\beta_p=10$ as the comparison case. We point that the ICM is observed to be weakly magnetized with a characteristic value of $\beta_p\sim 100$ \citep[see, e.g.,][]{ryu2008,porter2015}.

In turbulent flows, the magnetic field is generated via the so-called small-scale turbulent dynamo, and the magnetic energy density approaches equipartition with the kinetic energy density \citep[see, e.g.,][]{cho2009,zrake2012}. Considering that turbulence is ubiquitous in jet flows as demonstrated in Paper II, we introduce the dynamo-amplified field strength, $B_{\rm turb}$, defined as
\begin{equation}
\frac{B_{\rm turb}^{2}}{8\pi} \approx \mathcal{E}_{\rm turb},
\label{bturb}
\end{equation}
where $\mathcal{E}_{\rm turb}$ is the kinetic energy density of turbulent flow. Here, $\mathcal{E}_{\rm turb} = \Gamma_{\rm turb}(\Gamma_{\rm turb}-1)\rho c^2$ and $\Gamma_{\rm turb}= (1-(u_{\rm turb}/c)^2)^{-1/2}$; the turbulent flow velocity, $u_{\rm turb}$, is estimated by filtering the large-scale flow motions as described in Paper II (see also Section \ref{s3.3}).

Furthermore, numerous shocks arise in jet flows (see Figure \ref{f4}(a) below), and the magnetic field is expected to be amplified via both Bell's resonant and nonresonant CR streaming instabilities near the shocks \citep[see, e.g.,][]{bell2004,caprioli2014a,caprioli2014b}. So we calculate $B_{\rm Bell}$ at ``shock zones'' (grid zones with shocks) in the simulated jet-induced flows, defined as 
\begin{equation}
\frac{B_{\rm Bell}^{2}}{8\pi} \approx \frac{3}{2}\frac{u_s}{c}P_{\rm CR}.
\label{bmodel2}
\end{equation}
In our model, the CR pressure is approximated as $P_{\rm CR} \approx 0.1\rho_1u_s^2$ with the preshock density $\rho_1$ and the shock speed $u_s$, which reflects the DSA simulation results for non-relativistic shocks \citep{caprioli2014a}.

We then take a practical, yet physically motivated approach, in which the highest estimate is chosen among the three model values:
\begin{equation}
B = {\rm max}(B_{p},B_{\rm turb},B_{\rm Bell}).
\label{bmodel}
\end{equation}
It is done in the fluid frame, and this ``comoving'' magnetic field strength is used in the calculation of the particle MFP and the mean acceleration timescales described below. 

Panels (a) and (b) of Figure \ref{f3} show the 2D slice images of the ``comoving'' magnetic field in the fluid frame, $B$, and the ``observing'' magnetic field in the simulation frame (i.e., observer frame), $B_{\rm obs}$, for the Q46-$\eta5$-S model with $\beta_p=100$. To obtain $B_{\rm obs}$, $B$ has to be Lorentz-transformed, since the magnetic field is a frame-dependent quantity. Without the vector information of the magnetic field, however, we approximate the magnetic field strength in the observer frame as $B_{\rm obs}\approx \Gamma_f B$, where $\Gamma_f$ is the fluid Lorentz factor calculated with the fluid speed, $u$, in the simulation frame.

In the background ICM, the flow is static, and hence $B= B_{\rm obs}=B_{p}$; with $\beta_p=100$, the ICM has $B_{\rm obs}\approx 1\muG$ in our setup, which is the typical magnetic field strength observed in the ICM \citep[e.g.,][]{carilli2002,govoni2004}. Our adopted model results in $B_{\rm obs}\sim 10-100 \muG$ in the backflow and the jet-spine flow, which is in a good agreement with the magnetic field strength inferred from X-ray and radio observations of radio galaxies \citep[e.g.,][]{begelman1984,kataoka2005,anderson2022}.

In Figure \ref{f3}(c), {the probability distribution functions (PDFs)} of the comoving magnetic field $B$ with $\beta_p= 100$ in the jet-spine flow (red) and the backflow (blue) are plotted for the three ``S'' models (thick lines); the PDFs of $B$ with $\beta_p=10$ for the Q46-$\eta5$-S model are also presented (thin solid lines). The peaks of the PDFs represent relatively quiet zones where $B\approx B_p$, while the broad distributions mainly include dynamically active zones where $B\approx B_{\rm turb}$. With $\beta_p=100$, $B_{\rm turb}$ is likely to be larger than $B_p$ in the turbulent jet-spine flow and backflow, while $B_{\rm Bell}$ is dominant at shock zones. If we adopt $\beta_p=10$, however, $B_p$ could be larger than $B_{\rm turb}$ even in some zones of turbulent flows. The PDFs for the three ``S'' models with $\beta_p=100$ show that the magnetic field is stronger in more powerful jets, since flows with higher pressure $P$ and higher $\Gamma_{\rm turb}$ are produced.

Figure \ref{f2}(c) shows the time evolution of the volume-averaged, comoving magnetic field strength in the cocoon, $\langle B(t) \rangle$, estimated with $\beta_p= 100$ for some models considered. As expected, $\langle B(t) \rangle$ is larger in the jets with higher $Q_j$. Also, $\langle B(t) \rangle$ decreases in time due to both the lateral and radial expansions of the cocoon. Although nonlinear structures are better resolved in high-resolution simulations, we find that $\langle B(t) \rangle$ is not very sensitive to the resolution and is almost identical in the Q46-$\eta5$-S and Q46-$\eta5$-H models.

We note that our modeling of the magnetic field, although it is physically motivated, is still somewhat arbitrary. The resulting $B$ affects the MFP and hence the particle acceleration, as shown in the next section. In Section \ref{s5}, we compare the energy spectra of UHECRs in the Q46-$\eta5$-H jet for a specific scattering model with $\beta_p=100$ and $10$. Stronger magnetic field results in the spectrum that shifts a little to higher energies. However, we find that once $B$ is in the range observed in radio jets, the difference due to different magnetic field modelings would not be substantial.

\section{Particle Acceleration Physics}\label{s3}

Figure \ref{f4} presents the 2D images exhibiting the nonlinear flow dynamics for the jet in the stratified ICM in the high-resolution model Q46-$\eta5$-H at $t=t_{\rm end}$: (1) shocks, such as recollimation and turbulent shocks in the jet-spine flow, and numerous turbulent shocks in the backflow that are mostly non-relativistic (Figure \ref{f4}[a]), (2) turbulence in the jet-spine flow and the backflow (Figure \ref{f4}[b]), and (3) relativistic velocity shear along the interface between the jet-spine and the cocoon, and non-relativistic shear along the interface between the cocoon and the shocked ICM (Figure \ref{f4}[c]). These are overall similar to those for the jet models in the uniform medium presented in Paper II. With these nonlinear flows, it is expected that CRs are accelerated through {diffusive shock acceleration (DSA), turbulence shear acceleration (TSA), and relativistic shear acceleration (RSA)} in radio jets, as noted in the introduction; RSA may be further subdivided into gradual shear acceleration (GSA), and non-gradual shear acceleration (nGSA). Below, we briefly review these acceleration processes. We also define the corresponding acceleration timescales, $t_{\rm DSA}$, $t_{\rm TSA}$, $t_{\rm GSA}$, and $t_{\rm nGSA}$, to be used to assess the relative importance among the acceleration processes.

\subsection{Scattering of Particles}\label{s3.1}

Scattering of particles off underlying MHD fluctuations is the key element that governs the particle transport in both the spatial and momentum spaces, acceleration, confinement, and escape from the system. The important measure is the gyroradius of particles with energy $E$, which is given as
\begin{equation}
r_g \approx \frac{1.1~{\rm kpc}}{Z_i}\left(\frac{E}{ 1~{\rm EeV}}\right)\left(\frac{B}{1~\muG}\right)^{-1},
\label{rg}
\end{equation}
where $Z_i$ is the charge of CR nuclei. The maximum energy derived from the confinement condition that $r_g$ is equal to the radius of the acceleration system, $\mathcal{R}$, is given as
\begin{equation}
E_{H,\mathcal{R}} \approx 0.9~{\rm EeV}\cdot {Z_{i}}\left(\frac{\it{B}}{1~\mu\rm{G}}\right) \left( \frac{\mathcal{R}}{1~\rm{kpc}}\right).
\label{Ehillas}
\end{equation}
This geometrical condition is known as the ``Hillas criterion'', and $E_{H,\mathcal{R}}$ is referred as the ``Hillas energy'' \citep{hillas1984}. It provides a rough estimate of the energy with which particles are confined before escaping from the system. 

In general, the MFP of CRs, $\lambda_f (E)$, is thought to be momentum or energy-dependent. In a magnetized, strongly turbulent, collisionless plasma, the diffusion of particles across the magnetic field is often conjectured to follow the Bohm diffusion, leading to $\lambda_f (E)\sim r_g$ \citep{bohm1949}. It is known that at shocks, the self-generated magnetic fluctuations via various microinstabilities, such as resonant and nonresonant CR streaming instabilities, can be described by the Bohm limit, and hence $\lambda_f (E)\propto E$ \citep[e.g.,][]{caprioli2014a}. On the other hand, it is argued that in Kolmogorov-type turbulence, resonant scattering results in $\lambda_f (E)\propto E^{1/3}$ \citep[e.g.,][]{stawarz2008}, while on scales larger than the coherence length of turbulence, nonresonant scattering might result in $\lambda_f (E)\propto E^2$ \citep[e.g.,][]{sironi2013}. Hence, for instance, in \citet{kimura2018}, the MFP was assumed to be scaled with energy as $\lambda_f (E)\propto E^{1/3}$ on small scales and $\lambda_f (E)\propto E^2$ on large scales in the cocoon.

We here adopt a simple prescription for the mean-free-path (MFP),
\begin{equation}
\lambda_f (E) = \left(\frac{E}{E_{H,L_0}}\right)^{\delta}L_0,
\label{mf}
\end{equation}
where $L_0\sim r_j$ is the coherence length of turbulence in our jet simulations (see Paper II), and $E_{H,L_0}$ is the Hillas energy at the coherence length. Then, the mean scattering time is given as $\tau(p)\approx\lambda_f/c\propto p^\delta$, where $p\approx E/c$ is the momentum of CRs. Here, both $\lambda_f(E)$ and $\tau(p)$ are defined in the local fluid frame (i.e., scattering frame).

In the fiducial model, we adopt $\delta=1/3$ for $E<E_{H,L_0}$ and $\delta=1$ for $E>E_{H,L_0}$ in both the jet-spine flow and the backflow. If the particle is located in a shock zone, however, $\delta=1$ is assigned, regardless of its energy. In order to explore the dependence of the acceleration of highest energy CRs on particle scattering (see Section \ref{s5.3} for the discussion), we consider two additional models specified in Table \ref{t2}. In Model A, resonant scattering in Kolmogorov-type turbulence ($\delta=1/3$) is assumed for high-energy particles; in Model B, nonresonant scattering ($\delta=2$) is assumed for high-energy particles. Note that Model B is closest to, but slightly different from, that of \citet{kimura2018}, in which the Bohm-type diffusion is adopted in the jet-spine flow. In their simple geometrical setup, the jet-cocoon system consists of a upward-moving jet-spine flow and a laterally-expanding cocoon, so they focused mainly on RSA and did not include DSA.

\begin{deluxetable}{ll}[t]
\tablecaption{MFP Models \label{t2}}
\tabletypesize{\small}
\tablecolumns{2}
\tablenum{2}
\tablewidth{0pt}
\tablehead{
\colhead{mean-free-path (MFP):} &
\colhead{$\lambda_f=(E/E_{H,L_0})^{\delta} L_0$}}
\startdata
Bohm scattering:& $\delta=1$\\
Kolmogorov scattering:& $\delta=1/3$\\
nonresonant scattering:& $\delta=2$ \\
\hline
fiducial: & $E<E_{H,L_0}$: $\delta=1/3$, $\delta=1$ (at shocks) \\
   & $E>E_{H,L_0}$: $\delta=1$ \\
\hline
Model A: & $E<E_{H,L_0}$: $\delta=1/3$, $\delta=1$ (at shocks) \\
   & $E>E_{H,L_0}$: $\delta=1/3$ \\
\hline
Model B: & $E<E_{H,L_0}$: $\delta=1/3$, $\delta=1$ (at shocks) \\
   & $E>E_{H,L_0}$: $\delta=2$ \\
\enddata
\end{deluxetable}

\subsection{Diffusive Shock Acceleration {(DSA)}}\label{s3.2}

\citet{matthews2019} suggested that non-relativistic or mildly relativistic shocks in the lobe can effectively accelerate UHECRs via DSA, while relativistic shocks such as terminal shocks and recollimation shocks would be less efficient as particle accelerators \citep[e.g.][]{bell2018}. The maximum energy of particles achievable at astrophysical shocks can be estimated from the condition that the diffusion length of UHECRs in the Bohm limit, $l_{\rm diff}\sim\lambda_f(c/u_s)$, should be smaller than the shock size, $r_s$:
\begin{equation}
E_{\rm shock,max} \approx 0.9~{\rm EeV}\cdot {Z_{i}}\left( \frac{\it{B}}{1~\mu \rm{G}}\right) \left( \frac{\it{u_s}}{\it{c}}\right) \left(\frac{\it{r_s}}{1~\rm{kpc}}\right),
\label{hillas}
\end{equation}
where $u_s$ is the shock velocity. If $r_s\sim 1-10$ kpc and $B\sim 10-100~\muG$ in the lobe, radio jets could be potential sources of UHECRs of up to $E\sim10^{20}$ eV.

In the test particle regime of DSA, the energy spectrum of CRs accelerated by non-relativistic shocks takes a simple power-law form, $d\mathcal{N}/dE \propto E^{-\sigma}$, where the slope, $\sigma=(\chi+2)/(\chi-1)$, is determined solely by the compression ratio across the shock jump, $\chi=\rho_2/\rho_1$ \citep[e.g.,][]{bell1978,drury1983}. For instance, for strong non-relativistic shocks with $M_s\gg 1$, $\chi=4$ and $\sigma=2$.

The acceleration physics at relativistic shocks is more complex, and they depend on the shock speed and the magnetic field obliquity, as well as the particle scattering laws, among other parameters, in addition to the shock compression \citep[see][and references therein]{sironi2015}. In the test particle regime for ultra-relativistic shocks, the power-law slope approaches to $\sigma\approx 2.2$, and hence the energy spectrum tends to be steeper than that of strong non-relativistic shocks. \citep[e.g.,][]{kirk2000,keshet2005,ellison2013}.

{On the other hand, since numerous shocks form in the turbulent, jet-induced flows, DSA by multiple shocks is highly pertinent in this situation. Reacceleration by multiple, nonrelativistic shocks is known to flatten the DSA power-law spectrum to $\propto p^{-3}$, independent of the shock compression ratio \citep[e.g.][]{melrose1993,casse2003}. If the effects of other acceleration processes such as TSA and GSA (see below) are included as well, the energy spectrum of such multi-shock accelerated particles could be even flatter than $E^{-1}$.}

The mean DSA timescale for CR protons at non-relativistic shocks is given as
\begin{equation}
t_{\rm DSA} \approx 3.52\times10^3 {\rm yrs}~ {\frac{\chi(\chi+1)}{\chi-1}}\left(\frac{u_s}{c}\right)^{-2}\left(\frac{E}{1~\rm{EeV}}\right)\left(\frac{B}{1~\muG }\right)^{\rm{-1}},
\label{tshock}
\end{equation}
where a Bohm diffusion coefficient, $\kappa_B\approx (3.13\times 10^{22}~{\rm cm^2~ s^{-1}})(B/1\muG)^{-1}({p}/m_pc)$ is adopted \citep{drury1983}. Here, $m_p$ is the proton mass. As discussed in detail in Paper II, we identify ``shock zones'' in the simulated jet-induced flows, as shown in Figure \ref{f4}(a). The shock properties, such as $M_s$ and $u_s$, are calculated from the simulation data and used to estimate $t_{\rm DSA}$.

\subsection{Turbulent Shear Acceleration {(TSA)}}\label{s3.3}

The velocity shear appearing at the interfaces between the jet-spine flow and the backflow and between the backflow and the shocked ICM (Figure \ref{f4}[c]) is subject to Kelvin-Helmholtz instability, resulting in turbulence all over the jet-induced structures. In Paper II, filtering out the bulk jet motions on scales larger than the characteristic scale of the jet, $L_0\sim r_j$, we found that the jet-spine flow and the backflow exhibit the turbulence of Kolmogorov power-law, $\propto k^{-5/3}$, where the solenoidal mode dominates over the compressive mode. Hence, the so-called ``turbulent shear acceleration (TSA)'' by incompressible turbulence is expected to operate for particles whose MFP is smaller than $\sim r_j$ \citep[e.g.,][]{ohira2013}. For particles with $\lambda_f(p) \gtrsim r_j$, on the other hand, the relativistic shear acceleration would become important, which will be discussed in the next subsection.

As described in Paper II, the turbulent component of flow motions is extracted as \citep{vazza2017}:
\begin{equation}
\mbox{\boldmath$u$}_{\rm turb}(\mbox{\boldmath$r$}) = \frac{\mbox{\boldmath$u$}(\mbox{\boldmath$r$})-\left<\mbox{\boldmath$u$}(\mbox{\boldmath$r$})\right>_{L_0}}{1-\mbox{\boldmath$u$}(\mbox{\boldmath$r$})\cdot \left<\mbox{\boldmath$u$}(\mbox{\boldmath$r$})\right>_{L_0}\big/c^2}, 
\label{vtub}
\end{equation}
where $\left<\mbox{\boldmath$u$}(\mbox{\boldmath$r$})\right>_{L_0}={\sum_i w_i \mbox{\boldmath$u$}_i}/{\sum_i w_i}$ is the mean of the flow velocity averaged over the cubic box of size $L_0$. Here, we use a simple weight function, $w_i=1$, and set $L_0=r_j$. This method cannot perfectly separate the turbulent component from the strong laminar component in the direction of the jet propagation. So assuming that the turbulent velocity is almost isotropic, i.e., $u_{\rm turb,\it x}\approx u_{\rm turb,\it y}\approx u_{\rm turb,\it z}$, the turbulence speed is approximated as
\begin{equation}
|u_{\rm turb}|\approx \left[\frac{3}{2}( u^2_{\rm turb,\it x}+u^2_{\rm turb,\it y})\right]^{1/2}.
\label{vtur}
\end{equation}
Figure \ref{f4}(b) shows the 2D slice image of $|u_{\rm turb}|$.

\citet{ohira2013} considered TSA in non-relativistic incompressible turbulence. He derived analytic solutions for the momentum diffusion coefficient, $D_{\rm TSA}$, and the acceleration timescale, $t_{\rm TSA}= {p}^2/D_{\rm TSA}$, when the turbulence is of Kolmogorov-type. We here adopt Equation (19) of \citet{ohira2013} as follows:
\begin{equation}
t_{\rm TSA} \approx 2.88\times10^4 {\rm yrs} \left(\frac{L_0/\Gamma}{1~{\rm kpc}}\right)^{2/3}\left(\frac{|u_{\rm turb}|}{c}\right)^{-2}\left(\frac{\lambda_f(E)}{1~{\rm kpc}}\right)^{1/3},
\label{tturb}
\end{equation}
where the relativistic length contraction effect is included in the $L_0$ term. The energy spectrum of CRs produced by TSA depends on $\lambda_f$ and the characteristics of turbulence. In general, it does not take a simple power-law form.

\subsection{Relativistic Shear Acceleration {(RSA)}}\label{s3.4}

Once shear, $\Omega_{\rm shear} = |{\partial u_z}/{\partial r}|$, develops, particles can be energized by encountering the velocity difference due to the shear, $\Delta u=\Omega_{\rm shear}\lambda_f$, as they are elastically scattered off MHD fluctuations frozen into the flow \citep[see, e.g.,][for a review]{rieger2019}. Particles with $\lambda_f(E) \lesssim \Delta r$ (the width of the shear layer) undergo the stochastic acceleration process inside the shear layer, which is called {\it gradual} shear acceleration (GSA). On the other hand, particles with $\lambda_f(E) > \Delta r$ experience the whole velocity discontinuity by crossing the entire shear layer on each scattering, and can undergo the so-called discrete or {\it non-gradual} shear acceleration (nGSA).

As shown in Figure \ref{f4}(c), $\Omega_{\rm shear}$ is the largest along the interface between the jet-spine flow and the backflow, and hence GSA and nGSA operate mostly across this interface. The shear along the interface, which is relativistic, $\Omega_{\rm shear}r_j/c \sim 1$, has the width of the order of the jet radius, $\Delta r \sim r_j$ (see also Paper II). So particles with $\lambda_f(E) \lesssim r_j$ are expected to be accelerated by GSA, while a fraction of high energy particles with $\lambda_f(E) > r_j$ are accelerated via nGSA.

For relativistic GSA, the acceleration timescale was estimated, for instance, by \citet{webb2018}. We adopt the timescale in their Equation (21),
\begin{eqnarray}
t_{\rm GSA} & = &\frac{15}{(4+\delta)\Gamma_z^4 \Omega_{\rm shear}^2\tau({p})} \nonumber \\ 
\approx 4.90 & \times & 10^4 {\rm yrs} \frac{1}{(4+\delta)\Gamma_z^4}\left(\frac{\Omega_{\rm shear}}{c/r_j}\right)^{-2}\left(\frac{\lambda_f(E)}{1~{\rm kpc}}\right)^{-1},
\label{tshear}
\end{eqnarray}
where $\Gamma_z=(1-(u_z/c)^2)^{-1/2}$ and $\lambda_f(E)\propto E^\delta$ is used. Note that particles with longer $\lambda_f(E)$ experience larger velocity differences in the shear flow, so $t_{\rm GSA}$ is inversely proportional to the particle MFP.

Particles injected to the nGSA process with $\lambda_f(E) > \Delta r$ gain energy, on average $\langle \Delta E/E \rangle \sim (\Gamma_\Delta-1)$ per each crossing of the velocity discontinuity, $\Delta u$, across the shear layer, where $\Gamma_\Delta= (1-(\Delta u/c)^2)^{-1/2}$ \citep{rieger2004}. The mean energy gain per each cycle of crossing and recrossing the shear layer is given as $\langle \Delta E/E \rangle \sim (\Gamma_\Delta^2-1)$, if the particle momentum distribution is almost isotropic. However, this could be an overestimation since the velocity anisotropy could be substantial in relativistic shear flows. We take the mean acceleration timescale per cycle given in Equation (1) of \citet{kimura2018} only as an approximate measure:
\begin{equation}
t_{\rm nGSA} \sim \zeta \frac{\lambda_f(p)}{c \Gamma_z^2\beta_z^2}, \label{tngsa}
\end{equation}
where $\beta_z=u_z/c$ and $\zeta \sim 1$ is a numerical factor that mainly depends on the anisotropy of the particle distribution.

\citet{ostrowski1998} presented the analytic solution for the momentum distribution function, $f(p)$, for nGSA at a tangential discontinuity in the case of continuous mono-energetic injection. If there is no intrinsic scale in the system (such as the jet radius or the escape boundary size) and if particles are scattered with the mean scattering time of $\tau(p)\propto p^\delta$, the accelerated spectrum can be represented by $f( p) \propto p^{-3+\delta}$ (see their Equation [B2]), resulting in the energy spectrum of $d\mathcal{N}/dE = 4\pi p^2 f(p) \propto E^{-1+\delta}$. For the Bohm diffusion ($\delta = 1$), $d\mathcal{N}/dE\propto E^0$, which is much flatter than the canonical DSA power-law $d\mathcal{N}/dE\propto E^{-2}$ for strong non-relativistic shocks. 

\citet{kimura2018} performed Monte Carlo simulations for a mildly relativistic jet of $\Gamma_j\approx 1.4$, represented by the simplified jet-cocoon system in a cylindrical configuration. Seed galactic CRs are energized through large-angle scatterings in a manner of random walks. Adopting various scattering prescriptions for $\lambda_f(E)$ and a uniform magnetic field in the cocoon and the jet-spine flow, they found that nGSA produces a power-law spectrum of $d\mathcal{N}/dE\propto E^{-1}-E^{0}$ for escaping CRs. The high-energy end of the spectrum above the cutoff energy decreases more gradually than the exponential.

As mentioned in the introduction, \citet{caprioli2015} suggested the ``espresso'' scenario of particle acceleration, which is conceptually similar to nGSA being described here. If a particle experiences a ``one-shoot boost'' by crossing and recrossing the shear layer in an ultra-relativistic jet with $\Gamma_j\gg 1$, the energy is enhanced by a factor of $\sim \Gamma_j^2$ per cycle. In later studies, \citet{mbarek2019,mbarek2021} performed Monte Carlo simulations, in which seed CRs are injected into the self-consistent jet configuration from MHD simulations and their trajectories are followed. They found that particles could be accelerated to become UHECRs of $E\gtrsim10^{20}$ eV via one or two espresso shots.

\begin{figure}[t] 
\centering
\vskip 0.1 cm
\includegraphics[width=1\linewidth]{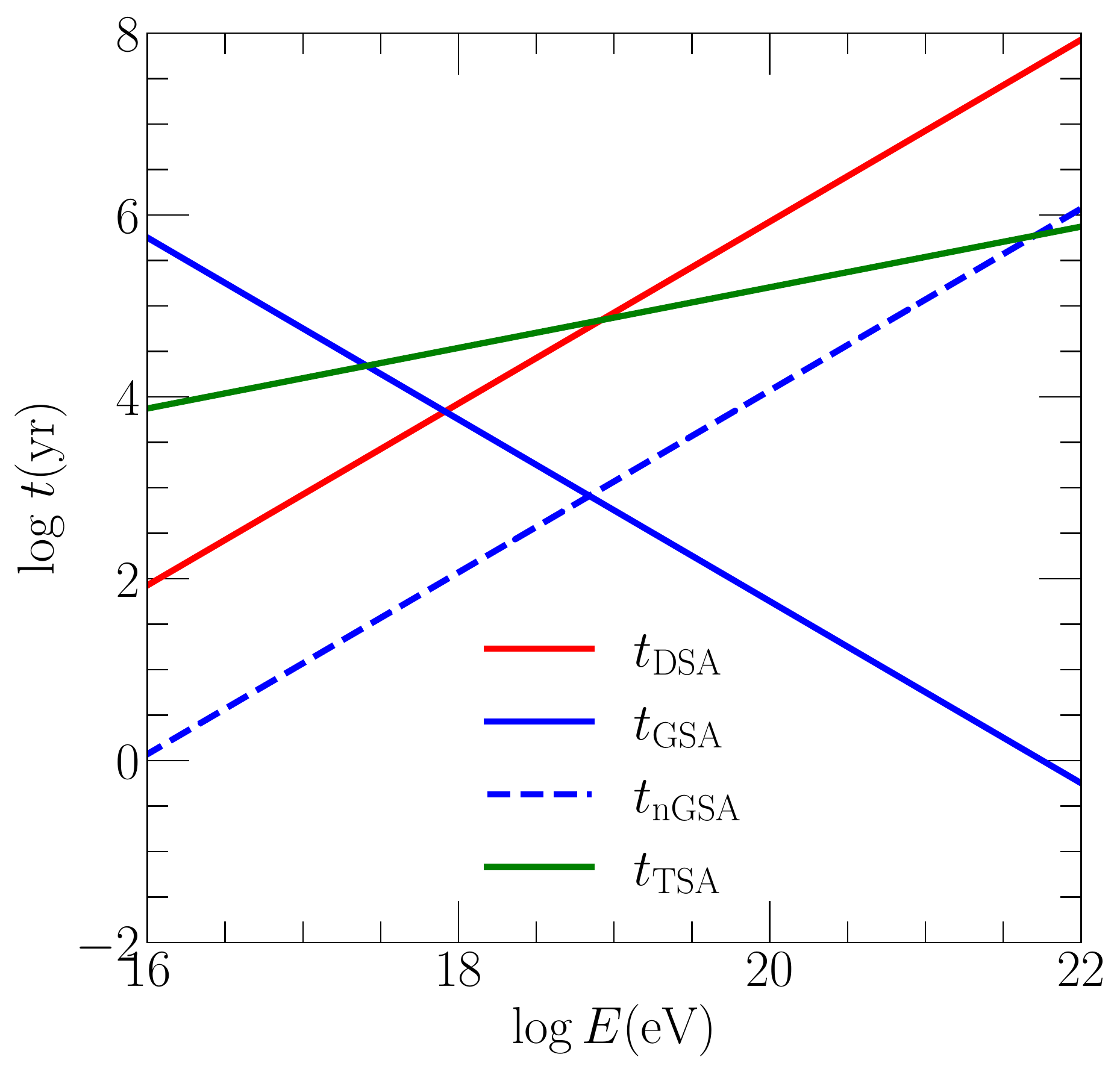}
\vskip -0.1 cm
\caption{Acceleration timescales for different processes: $t_{\rm DSA}$ (red), $t_{\rm TSA}$ (green), $t_{\rm GSA}$ (blue solid), and $t_{\rm nGSA}$ (blue dashed) given in Equations (\ref{tshock}), (\ref{tturb}), (\ref{tshear}), and (\ref{tngsa}), respectively. For illustrative purpose, we assume the following parameters: $\chi= 4$, $u_s/c \sim 0.5$, $B \sim 10~\muG$, $L_0\sim1$ kpc, $\Omega_{\rm shear} \sim 1~c/r_j$, $\Gamma\sim\Gamma_z \sim 2$, $|u_{\rm turb}|/c \sim 0.5$, and $\zeta\sim1$. For the MFP, $\lambda_f(p)$ in Equation (\ref{mf}) with $\delta=1$ is used.}
\label{f5}
\end{figure}

\subsection{Comparison of Acceleration Timescales}\label{s3.5}

To assess the relative importance of different acceleration processes, we compare the acceleration timescales given in Equations (\ref{tshock}), (\ref{tturb}), (\ref{tshear}), and (\ref{tngsa}) in Figure \ref{f5}. The adopted characteristic parameters are specified in the figure caption. For particles with $E\lesssim 10^{18}$ eV, $t_{\rm DSA}$ is the shorter than $t_{\rm TSA}$ and $t_{\rm GSA}$, and DSA would be the dominant acceleration process. For higher energy particles with $E\gtrsim 10^{18}$ eV, GSA would become important. As noted above, only a small fraction of particles, which are energized via other processes to have $\lambda_f(E) > r_j$, could cross the entire shear layer and are injected to the nGSA process. Hence, although $t_{\rm nGSA}$ is shortest even at low energies, nGSA operates only for the highest energy CRs. TSA, on the other hand, would be only marginally important around $E\sim 10^{18}$ eV.

{If the simple condition $t_{\rm DSA}\approx t_{\rm GSA}$ is adopted, the transition energy above which GSA becomes faster than DSA is roughly $E_{\rm trans}\approx 4~{\rm EeV}$ $(\langle u_s\rangle /c)(\langle B \rangle /1\muG)\langle \Gamma_j \rangle^{-2}(\langle \Omega_{\rm shear}\rangle r_j/c)^{-1}$. For the jet models considered here, the volume-averaged quantities range approximately as follows: $(\langle u_s\rangle /c)\sim 0.3-0.5$, $\langle \Gamma_j \rangle\sim 1.3 - 5$, $(\langle \Omega_{\rm shear}\rangle r_j/c)\sim 1-1.5$, $\langle B \rangle\sim 3-15\muG$ (see also Paper II). All of them increase slowly with $Q_j$, resulting in $E_{\rm trans}\sim$ EeV. In Section \ref{s5.1}, we will show that indeed, at low energies, DSA seems to be important, while at high energies, GSA and nGSA are dominant, in MC simulations.}

\subsection{Maximum Energy of Accelerated Particles}\label{s3.6}

In the early development stages of radio jets, the maximum energy, $E_{\rm max}$, that CRs can achieve is limited by the age. In later stages, $E_{\rm max}$ is expected to be limited by the size of the cocoon where CRs are confined. Since the smallest acceleration timescales in Figure \ref{f5} are much shorter than the duration of our simulations, $t_{\rm end}\approx 10^6-10^7$ yrs (see Table \ref{t1}), the size-limited $E_{\rm max}$ would be relevant.

Most CRs are expected to escape diffusively from the cocoon. So the size-limited $E_{\rm max}$ can be estimated from the condition that $\lambda_f(E)$ is equal to the radius of the cocoon. With the lateral width of the cocoon, $\mathcal{W}$, shown in Figure \ref{f2}(b), $\lambda_f(E)\sim\mathcal{W}/2$ gives
\begin{eqnarray}
E_{\rm max} & \approx & E_{H,L_0} \left(\frac{\mathcal{W}}{2L_0}\right) \nonumber \\
& \approx & 0.9~{\rm EeV}\cdot Z_i \left(\frac{B}{1~\muG}\right)\left(\frac{r_j}{1~{\rm kpc}}\right) \left(\frac{\mathcal{W}}{2r_j}\right), \label{Emax}
\end{eqnarray}
where again $L_0=r_j$ is used. For $\langle B \rangle\mathcal{W}\sim200$ (see Figure \ref{f2}[d]), $E_{\rm max}$ is estimated to be $\sim 200$ EeV for protons. Note that the above size-limited $E_{\rm max}$ does not directly depend on $\Gamma_j$; hence, $E_{\rm max}$ could be lower even in more powerful jet models with higher $\Gamma_j$, if $\mathcal{W}$ is smaller.

On the other hand, the length of the cocoon, $\mathcal{L}$, is greater than $\mathcal{W}$, as shown in Figure \ref{f2}(a). So CRs even with $E\gtrsim E_{\rm max}$ could be confined in the cocoon, and they may escape through the longitudinal direction. In addition, a small fraction of CRs can be boosted to much higher energies via nGSA, and their escape may not be described as a diffusive process. In Section \ref{s5.2}, we will discuss how such anisotropic escape affects the high-energy end of the spectrum of UHECRs.

\section{Monte Carlo Simulations}\label{s4}

\subsection{Recipes for Monte Carlo Simulations}\label{s4.1}

To track the acceleration of CRs in simulated jet-induced flows, we perform {Monte Carlo (MC}) simulations, where the trajectories of CRs are integrated, according to the following recipes:
(1) In the jet simulations described in Section \ref{s2}, the snapshot data of jet flow quantities are stored with a time interval of $\Delta t_s=(1/3)t_{\rm{cross}}$.
(2) At every $\Delta t_s$, 200 seed CRs are injected from the jet nozzle with energy spectrum, $d\mathcal{N}_{\rm inj}/dE_{\rm inj}\propto E_{\rm inj}^{-2.7}$, for $E_{\rm inj}=(0.01-1)$ PeV. Note that $-2.7$ is the slope of the Galactic CR energy spectrum \citep[e.g.,][]{thoudam2016}.
(3) CR particles are assumed to be scattered elastically off small-scale MHD fluctuations that are frozen in the background flow in the ``restricted'' random walk scheme (see Section \ref{s4.2}).
(4) Large angle scattering is imposed by hand, adopting the prescription for the energy-dependent MFP, $\lambda_f(E)$, in Equation (\ref{mf}) with the magnetic field model described in Section \ref{s2.4}.
(5) The probability of particle displacement at each scattering is assumed to obey the exponential distribution with $\lambda_f(E)$.
(6) At every elastic scattering event, the Lorentz transformation from the local fluid frame to the simulation frame is performed, which results in the energy change, $\Delta E$ (either positive or negative).
(7) The energy evolution of CR particles is calculated along the evolution of jet-induced flows using the snapshot data.
(8) The information on the particles escaping from the computational box is stored and used to analyze the properties of UHECRs, such as the energy spectrum.

Seed CRs with $E_{\rm inj}$ have the gyroradii that are much smaller than the grid size, i.e., $r_g\ll\Delta x =0.1-0.33$ kpc (see the 10th column of Table \ref{t1}), and hence go through scatterings within a grid zone. For the calculation of the energy change due to such subgrid scatterings, we employ the variation of the 3D fluid velocity inside a grid cell, which is approximated using trilinear interpolation with the fluid velocities along the direction of particle trajectory. These subgrid scatterings would be applied mostly to CRs of $E<10^{18}$ eV (see Equation [\ref{Ehillas}]).

\subsection{Restricted Random Walk}\label{s4.2}

In the calculation of conventional random walk transport, the scattering angle, $(\delta\mu,\delta\phi)$, is chosen randomly in the ranges of $-1\le \delta \mu \le +1$ and $0\le \delta\phi \le 2\pi$, where $\mu=\cos \theta$. In real jet flows, however, magnetic field fluctuations may not be large enough to scatter fully isotropically high-energy particles with $\lambda_f(E)\gtrsim L_0$. Thus, to mimic roughly pitch-angle diffusion without knowing the magnetic field configuration, we adopt a random walk scheme, in which the scattering angle with respect to the incident direction is chosen from the ``restricted'' range of $\delta\mu_{\rm max} \le \delta\mu\le +1$. In addition, we employ an energy-dependence, with which the scattering angle with respect to the incident direction is forced to be smaller than $\sim \pi[L_0/\lambda_f(E)]$. Without {\it a prior} knowledge of the turbulent nature of magnetic field fluctuations, it may provide a crude model to account for pitch-angle scattering in an energy-dependent way.

In our random walk scheme, the maximum value of scattering angle is modeled specifically as
\begin{equation}
\delta \theta_{\rm max} \approx \pi \cdot \min\left[\psi\left(\frac{ L_0}{\lambda_f}\right), 1\right]. 
\label{thetamax}
\end{equation}
Here, $\psi \lesssim 1$ is a free parameter that is devised to reflect the strength of magnetic field fluctuations. Again, $L_0\sim r_j$ is assumed for the coherence scale of turbulence. For high-energy particles with $\lambda_f(E)\gg L_0$, this prescription leads to forward-beamed scattering with $\delta \theta_{\rm max}\ll 1$. For low-energy particles with $\lambda_f(E)\ll L_0$, by contrast, it results in isotropic scattering with $\delta \theta_{\rm max}\approx \pi$. Adopting this scheme tends to reduce the energy gain of high-energy particles, especially, for nGSA near the jet-backflow interface. We take $\psi=1$ as the fiducial value, and also present the isotropic scattering case ($\psi\rightarrow\infty$) for the demonstration of model dependence.

\begin{figure*}[t]
\centering
\vskip 0.1 cm
\includegraphics[width=1\linewidth]{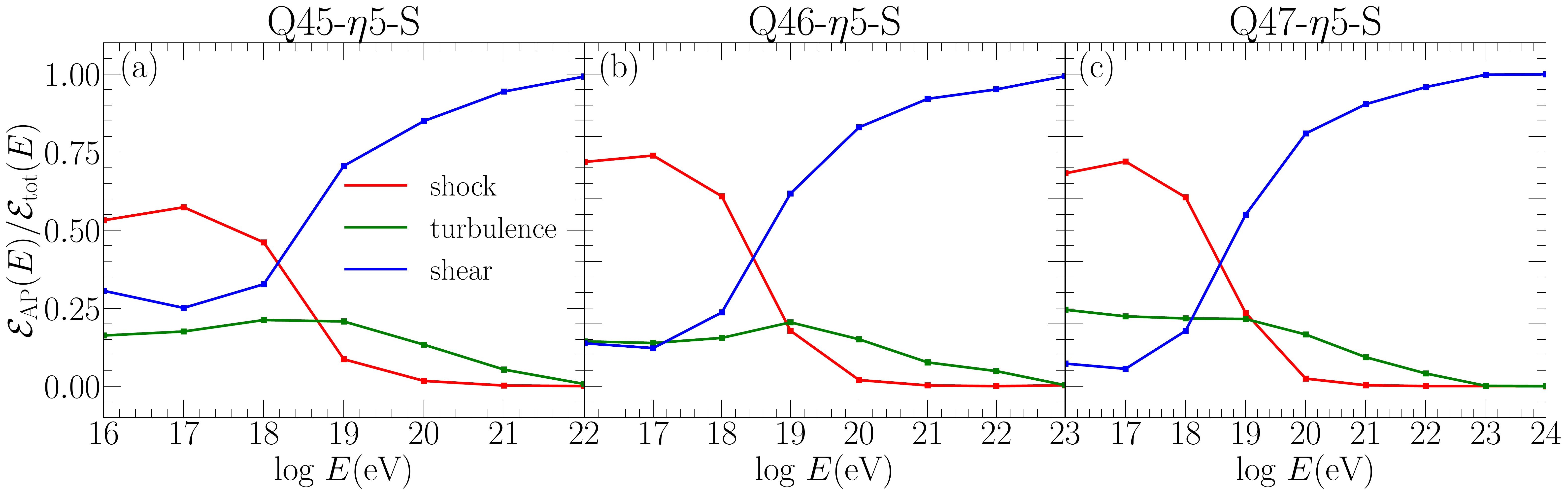}
\vskip -0.1 cm
\caption{Fraction of the cumulative energy gains due to different APs, $\mathscr{E}_{\rm AP}(E)/\mathscr{E}_{\rm tot}(E)$, as a function of the particle energy for the three ``S'' models, (a) Q45-$\eta$5-S, (b) Q46-$\eta$5-S, and (c) Q47-$\eta$5-S, at $t=t_{\rm{end}}$. Here, $\mathscr{E}_{\rm AP}(E)$ is the {\it weighted} sum, $\sum \xi_{\rm AP} \Delta E$, contributed by the given type of AP for all the escaping particles whose final energy lies in the logarithmic bin of $[\log E, \log E + d\log E]$. The cases of shock (red), turbulence (green), and shear (blue) accelerations are shown. The energy range in the abscissa is different in different models, because the highest energy reached is different in jets with different powers (see the next subsection).}
\label{f6}
\end{figure*}

\subsection{Primary Acceleration Process}\label{s4.3}

As illustrated in Figure \ref{f5}, in jet-induced flows, {diffusive shock acceleration (DSA) and turbulent shear acceleration (TSA) would be important for low-energy particles, while gradual shear acceleration (GSA) and non-gradual shear acceleration (nGSA)} would become more important for higher energy particles near the jet-backflow interface. In MC simulations, however, in each scattering, the change in the particle energy often involves a combination of the different processes.

In an effort to evaluate the relative importance of the acceleration processes, we estimate the acceleration timescales, $t_{\rm TSA}$ and $t_{\rm GSA}$, at each scattering. If the particle crosses a single or multiple shock zones in the displacement after scattering, $t_{\rm DSA}$ is also calculated. We then attempt to identify the {\it primary} acceleration process (PAP) by determining the shortest acceleration timescale among the two or three timescales. If DSA is chosen as PAP, the scattering event is tagged as ``shock"; if TSA is chosen, it is tagged as ``turbulence"; if GSA is chosen, it is tagged as ``shear". We do not include $t_{\rm nGSA}$ in the PAP selection, since only a very small fraction of high energy particles undergo nGSA. Through this crude evaluation, we will see that DSA is indeed the PAP for $E\lesssim 1$ EeV, while GSA becomes dominant above EeV, as presented in the next section.

\section{Results}\label{s5}

We here focus on CR protons ($Z_i=1$) escaping from the cocoon.

\begin{deluxetable}{ccccc}[t]
\tablecaption{Energy Gains via Different APs \label{t3}}
\tabletypesize{\small}
\tablecolumns{4}
\tablenum{3}
\tablewidth{0pt}
\tablehead{
\colhead{Model name} &
\colhead{$\frac{\tilde{\mathscr{E}}_{\rm{shock}}}{\tilde{\mathscr{E}}_{\rm{tot}}}$(\%)} &
\colhead{$\frac{\tilde{\mathscr{E}}_{\rm{turb}}}{\tilde{\mathscr{E}}_{\rm{tot}}}$(\%)} &
\colhead{$\frac{\tilde{\mathscr{E}}_{\rm{shear}}}{\tilde{\mathscr{E}}_{\rm{tot}}}$(\%)}
}
\startdata
Q45-$\eta5$-S &3.2&13.6&83.2 \\
Q46-$\eta5$-S &2.2&12.5&85.3\\
Q47-$\eta5$-S &1.9&12.2&85.9\\
\hline
\enddata
\end{deluxetable}

\subsection{Relative Contributions of Acceleration Processes}\label{s5.1}

\begin{figure*}[t] 
\centering
\vskip 0.1 cm
\includegraphics[width=1\linewidth]{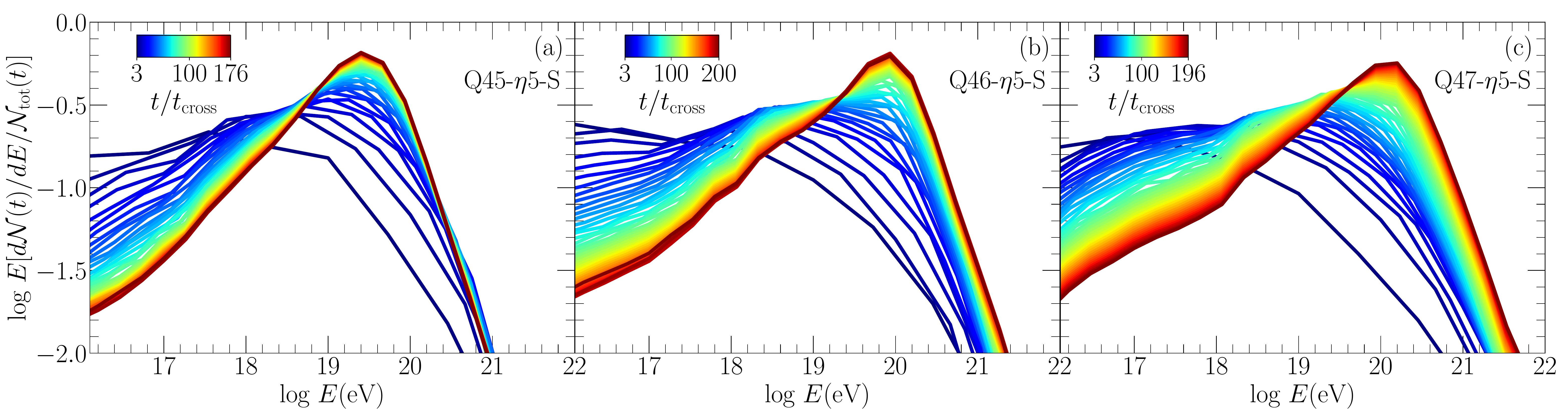}
\vskip -0.1 cm
\caption{(a) Time-integrated energy spectrum, $[E/\mathcal{N}_{\rm tot}(t)][d\mathcal{N}(t)/dE]$, of all particles escaping from the system up to a given time, $t$, for the three ``S'' models, (a) Q45-$\eta$5-S, (b) Q46-$\eta$5-S, and (c) Q47-$\eta$5-S. Here, $\mathcal{N}_{\rm tot} = \int (d\mathcal{N}/dE) dE$. The lines are color-coded from blue to brown, based on the time, $t/t_{\rm cross}$, during the period of $3~t_{\rm cross}-t_{\rm end}$.}
\label{f7}
\end{figure*}

We begin with discussions on the relative contributions of different acceleration processes (APs). For each scattering event, a fraction of $\Delta E$ is assigned to each AP, according to the weight function, $\xi_{\rm AP}=t_{\rm AP}^{-1}/\sum_{\rm AP}^{}{t_{\rm AP}^{-1}}$, where the summation includes the three APs as described in Section \ref{s4.3}. For all particles escaping from the system until $t=t_{\rm{end}}$, whose final energy lies in the logarithmic bin of $[\log E, \log E + d\log E]$, the contribution of $\xi_{\rm AP}\Delta E$ are summed to make the cumulative energy gains, $\mathscr{E}_{\rm AP}(E)$, for each AP. 

Figure \ref{f6} shows the fraction, $\mathscr{E}_{\rm AP}(E)/\mathscr{E}_{\rm tot}(E)$, for the three APs as a function of the particle energy $E$. Here, AP stands for ``shock" (red), ``turbulence" (green), and ``shear" (blue), and $\mathscr{E}_{\rm tot}(E)= \mathscr{E}_{\rm shock}(E)+ \mathscr{E}_{\rm turb}(E)+\mathscr{E}_{\rm shear}(E)$. While these fractions should be only rough estimates, the figure confirms once again that for $E\lesssim 1$ EeV, particles are energized mainly by DSA, whereas GSA/nGSA becomes increasingly important above EeV. TSA makes only a supplementary contribution.

{We note in Figure \ref{f6} that for $E\lesssim 1$ EeV, the ratio of DSA to GSA/nGSA contributions, $\mathscr{E}_{\rm shock}/\mathscr{E}_{\rm shear}$, is higher for the jet models with higher $Q_j$. This seems contradictory to the simple expectation that GSA and nGSA would become more significant with higher $Q_j$ (higher $\Gamma_j$). In fact, the mean-free-path (MFP) is given as $\lambda_f\propto (E/B)^\delta$ in Equation (\ref{mf}) and $B$ scales approximately as $\propto Q_j^{0.3}$ (see Figure \ref{f2}[c]) in our models, reducing the probability of GSA and nGSA across the shear layer, especially for low-energy particles. So the relative importance of the different APs illustrated in the figure would be regarded as being specific to the various models employed here.}

We next calculate the total energy gains due to different APs, $\tilde{\mathscr{E}}_{\rm AP}=\sum \mathscr{E}_{\rm AP}(E)$, summed over all the particle energies. Table \ref{t3} lists the fractions, $ \tilde{\mathscr{E}}_{\rm AP}/\tilde{\mathscr{E}}_{\rm tot}$, where $\tilde{\mathscr{E}}_{\rm tot}= \tilde{\mathscr{E}}_{\rm shock}+\tilde{\mathscr{E}}_{\rm turb}+ \tilde{\mathscr{E}}_{\rm shear}$. Overall, shear acceleration (including both GSA and nGSA) is the dominant process, contributing to $\sim85\%$ of the energization of UHECRs, while DSA and TSA together generate only $\sim15\%$ of the energy. A noted point is that although TSA is subdominant in the whole energy range, its total contribution is larger than DSA. In addition, the contributions of DSA and TSA tend to be a bit larger for less powerful jet models; this is a consequence of more extended cocoons filled with shocks and turbulence in less powerful jets.

\subsection{Energy Spectrum of Accelerated Particles}\label{s5.2}

We next present the energy spectrum of escaping particles. The Monte Carlo simulations for GSA/nGSA by \citet{ostrowski1998} and \citet{kimura2018} showed that {the energy spectrum behaves as $\propto E^{-1} - E^0$ below the ``break energy'', $E_{\rm break}$, while it could be approximated by another steeper power-law above $E_{\rm break}$, instead of an exponential cutoff\footnote{In general, an exponential cutoff is expected at the high-energy end of the size-limited spectrum. For instance, it was shown that the spectrum of ``shock-accelerated'' particles (escaping without energy losses) could be represented by the DSA power-law with an exponential cutoff at the high-energy end, as $d\mathcal{N}/dE \propto E^{-\sigma}[1+(E/E_{\rm max})] \exp[-C (E/E_{\rm max})]$ with $C=(\sigma -1)$, when the Bohm diffusion is adopted \citep{protheroe1999}.}. Here, $E_{\rm break}$ is introduced to designate the energy above which the spectrum rolls over from one power-law to another power-law, whereas \citet{ostrowski1998} and \citet{kimura2018} used different terms.}

\begin{figure*}
\centering
\vskip 0.1 cm
\includegraphics[width=1\linewidth]{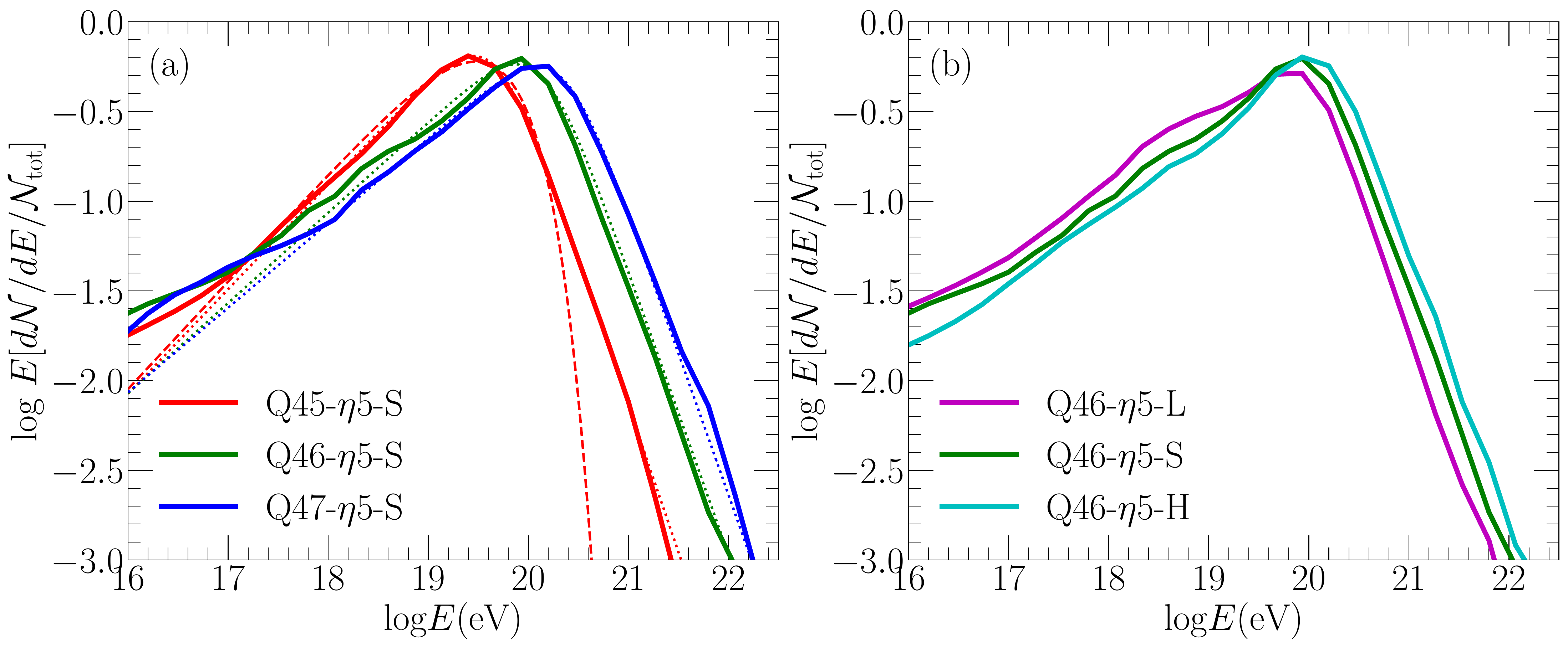}
\vskip -0.1 cm
\caption{Panel (a): Time-asymptotic energy spectra of all particles escaping from the system up to $t_{\rm end}$ for the three ``S'' models, Q45-$\eta5$-S (solid red), Q46-$\eta5$-S (solid green), and Q47-$\eta5$-S (solid blue). The fittings to Equation (\ref{spectfit}) with the fitting parameters presented in Table \ref{t4} are overlaid with dotted lines. The spectrum with an exponential cutoff above the break energy for Q45-$\eta5$-S is also shown for comparison (dashed red). Panel (b): Time-asymptotic energy spectra for the three Q46 models with different resolutions, Q46-$\eta5$-L (solid purple), Q46-$\eta5$-S (solid green), and Q46-$\eta5$-H (solid cyan). Note that the spectrum of Q46-$\eta5$-S are identical in the two panels.}
\label{f8}
\end{figure*}

As described in Section \ref{s4.1}, the underlying flow profile is updated and seed particles of $E_{\rm inj}=(0.01-1)$ PeV are injected at every $\Delta t_s=(1/3)t_{\rm{cross}}$. Hence, the energy spectrum should be time-dependent. We calculate the time-integrated, cumulative energy spectrum of {\it all} particles escaping from the system up to a given time $t$, $d\mathcal{N}(t)/dE$. Note that the total number of escaped particles, $\mathcal{N}_{\rm tot} = \int (d\mathcal{N}/dE) dE$, increases with time. As shown in Figures \ref{f5} and \ref{f6}, the dominant APs switch from DSA to GSA/nGSA at $E \sim 1$ EeV. Thus, as the seed population of $\propto E_{\rm inj}^{-2.7}$ is accelerated, the resulting spectrum is expected to flatten roughly from the DSA power-law of $E^{-2}$ at early stages to the GSA/nGSA power-law of $E^{-1}-E^{0}$.

Figure \ref{f7} shows the evolution of the time-integrated energy spectrum, $E[d\mathcal{N}(t)/dE]$, for the three ``S'' models. {Below $E_{\rm break}$,} the power-law portion continuously hardens over time, approaching $d\mathcal{N}/dE\propto E^{-0.5}$. At early epochs it is somewhat flatter than the DSA spectrum, because TSA and GSA also operate even at early stages. The time-asymptotic spectrum, $d\mathcal{N}/dE\propto E^{-0.5}$, is in agreement with previous works of \citet{ostrowski1998} and \citet{kimura2018}, indicating that GSA/nGSA would be the dominant energization process for particles around {$E_{\rm break}$} at late stages. {The break} shifts to higher energies with time during early stages, which is consistent with the age-limited maximum energy. It gradually approaches the size-limited maximum energy at late epochs (see below). For {$E> E_{\rm break}$,} the spectrum becomes steeper as time goes on, but it is still not as steep as the exponential drop even at $t_{\rm end}$. Overall, the time-integrated spectrum asymptotically saturates by the end of the simulations in all the models.

{Because the CR acceleration is expected to be size-limited at late stages, $E_{\rm max}$ in Equation (\ref{Emax}) is pertinent here, and $E_{\rm break}$ in the time-asymptotic spectrum would be similar to $E_{\rm max}$. The size-limited, maximum energy, $E_{\rm max}$, scales as $\propto B\mathcal{W}$. Figure \ref{f2} shows that the cocoon width, $\mathcal{W}$, increases in time, while the volume-averaged magnetic field strength, $\langle B \rangle$, decreases owing to the decreasing pressure in the laterally expanding cocoon. As a result, the value of $\langle B \rangle\mathcal{W}$ increases in time during the early stage and later approaches a time-asymptotic value for $t/t_{\rm cross}\gtrsim50$. The break of the energy spectrum in Figure \ref{f7} reflects these time-dependent behaviors of $E_{\rm max}(t)\propto \langle B \rangle\mathcal{W}$. In addition, the time-asymptotic value of $\langle B \rangle\mathcal{W}$ scales roughly as $\propto Q_j^{1/3}$ (Figure \ref{f2}[d]), and so does $E_{\rm max}$, for the model parameters under consideration here.} This is due to the fact that while $\mathcal{W}$ does not differ much in different jet models (see Figures \ref{f1} and \ref{f2}[b]), $\langle B\rangle$ is larger in higher power jets (Figure \ref{f2}[b]). 

As described in Section \ref{s2.3}, the jet power $Q_j$ is the primary parameter that determines the properties of jet-induced flows, such as the Lorentz factor of the jet flow, $\Gamma_j$, the cocoon's shape, and the associated nonlinear structures, which in turn govern the ensuing particle acceleration via DSA, TSA, and RSA. Figure \ref{f8}(a) shows the time-asymptotic energy spectrum, $E[d\mathcal{N}(t)/dE]$ at $t_{\rm end}$, for the models with different $Q_j$. The spectrum shifts to higher energies for higher $Q_j$, as expected. Otherwise, the overall shape of the spectrum is similar, except at the low-energy part of $E\lesssim 1$ EeV where the spectra differ due to different energization histories via DSA and TSA.

\begin{deluxetable}{ccccccc}[t]
\tablecaption{Fitting Parameters \label{t4}}
\tabletypesize{\small}
\tablecolumns{4}
\tablenum{4}
\tablewidth{0pt}
\tablehead{
\colhead{Model name} &
\colhead{$a$} &
\colhead{$b$} &
\colhead{$E_{\rm break}$(eV)} &
\colhead{$E_{\rm max}$(eV)$^1$}
}
\startdata
Q45-$\eta5$-S &0.59&1.64&4.5E19&7.0E19\\
Q46-$\eta5$-S &0.51&1.58&1.3E20&1.6E20\\
Q47-$\eta5$-S &0.47&1.60&2.2E20&2.8E20\\
\hline
\enddata
\tablenotetext{^1}{$E_{\rm max}$ is calculated with Equation (\ref{Emax}), {adopting the time-asymptotic values of $\langle B \rangle\mathcal{W}$.}} 
\end{deluxetable}

\begin{figure*}
\centering
\vskip 0.1 cm
\includegraphics[width=1\linewidth]{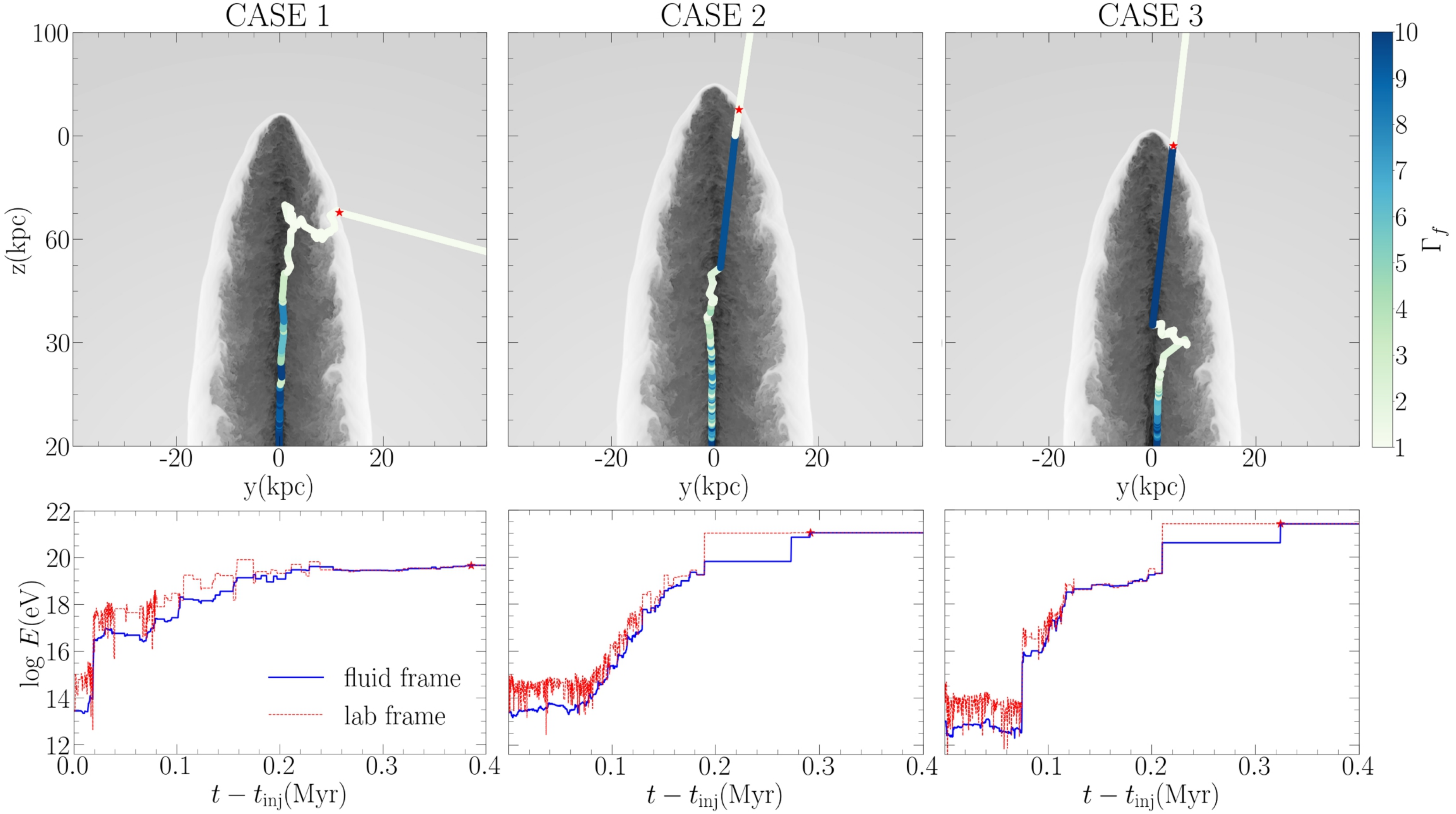}
\vskip -0.1 cm
\caption{{\it Top panels}: Trajectories of three sample particles since their injection into the jet flows of the Q46-$\eta5$-H model, illustrating CASE 1 (left), CASE 2 (middle), and CASE 3 (right). See the main text for the categorization of the cases. The trajectories are color-coded by the Lorentz factors of the fluid, $\Gamma_f$, at each scattering point, according to the color bar on the right. The red stars mark the points, where the particles exit the jet to the ICM. The background images show the 2D distributions of log $\rho$ at the exit time. {\it Bottom panels}: Energization history of the sample particles along the trajectories. The red and blue lines draw the energies of the particles at each scattering point in the simulation and fluid frames, respectively. At the times marked with the red stars, the energy in the fluid frame is adjusted to that of the ICM fluid frame (or the simulation frame).}
\label{f9}
\end{figure*}

To quantify the jet-power dependence, we attempt to fit the time-asymptotic spectrum in Figure \ref{f8}(a) to a functional form. As stated above, $d\mathcal{N}/dE$ is close to $\propto E^{-0.5}$ for {$E<E_{\rm break}$, while it drops roughly as another steeper power-law for $E>E_{\rm break}$.} Hence, instead of an exponential function, we employ the following double-power-law form:
\begin{equation}
\frac{Ed\mathcal{N}}{dE} \propto \left(\left(\frac{E}{E_{\rm{break}}}\right)^{-a}+\left(\frac{E}{E_{\rm{break}}}\right)^{b}\right)^{-1}
\label{spectfit}
\end{equation}
where $a$, $b$, and $E_{\rm{break}}$ are the fitting parameters. Table \ref{t4} lists these three fitting parameters and $E_{\rm max}$ in Equation (\ref{Emax}) estimated with the time-asymptotic values of $\langle B \rangle\mathcal{W}$. {Indeed, $E_{\rm{break}}$ is quite similar to $E_{\rm max}$.} In the figure, the fitted double-power-law spectra are overlaid with dotted lines. {We also plot $Ed\mathcal{N}/dE \propto (E/E_{\rm{break}})^a \exp(-E/E_{\rm{break}})$ for the Q45 model (dashed red) in the figure, to illustrate the difference between the power-low drop and the exponential cutoff beyond $E_{\rm{break}}$.}

We note that $E_{\rm{break}}$ shifts to higher values for higher $Q_j$. On the other hand, the power-law slopes, $a$ and $b$, show only a weak dependence on $Q_j$. On average, $a\sim0.5$, so $d\mathcal{N}/dE\propto E^{-0.5}$ below $E_{\rm break}$, as expected.

For $E>E_{\rm break}$, the power-law slope is, on average, close to $b \sim-1.6$, meaning $d\mathcal{N}/dE\propto E^{-2.6}$. To comprehend this slope, we examine the trajectories of particles and the ensuing energization in the MC simulations, focusing on RSA at the shear interface between the jet-spine flow and the backflow. As discussed in Section \ref{s3.5}, most of particles are incrementally accelerated via GSA and other processes, whereas a small fraction of high-energy particles with $\lambda_f(E) > r_j$ could be boosted in energy by a factor of $(\Gamma_{\Delta}^2-1)$ via nGSA if they cross and recross the shear interface. Typically, CRs with $E \gtrsim E_{\rm break}$ cross the shear interface more than once, before they escape from the jet to the ICM. In particular, the energization of highest energy particles seems to be governed by the experience of nGSA episodes. Based on this picture, we categorize escaping particles roughly into three cases, as illustrated in Figure \ref{f9}. In CASE 1, particles gain energy mainly via GSA and other processes and exit the cocoon to the ICM without experiencing a boost by nGSA. In CASE 2, after small incremental accelerations, particles cross into the jet-spine, and then cross out of the jet-spine into the cocoon, resulting in a $\sim\Gamma_{\Delta}^2$ boost via nGSA. They are confined in the cocoon, before they exit to the ICM. CASE 3 is the same as CASE 2, except that particles cross out of the jet-spine and exit directly to the ICM.

\begin{figure}
\centering
\vskip 0.1 cm
\includegraphics[width=1\linewidth]{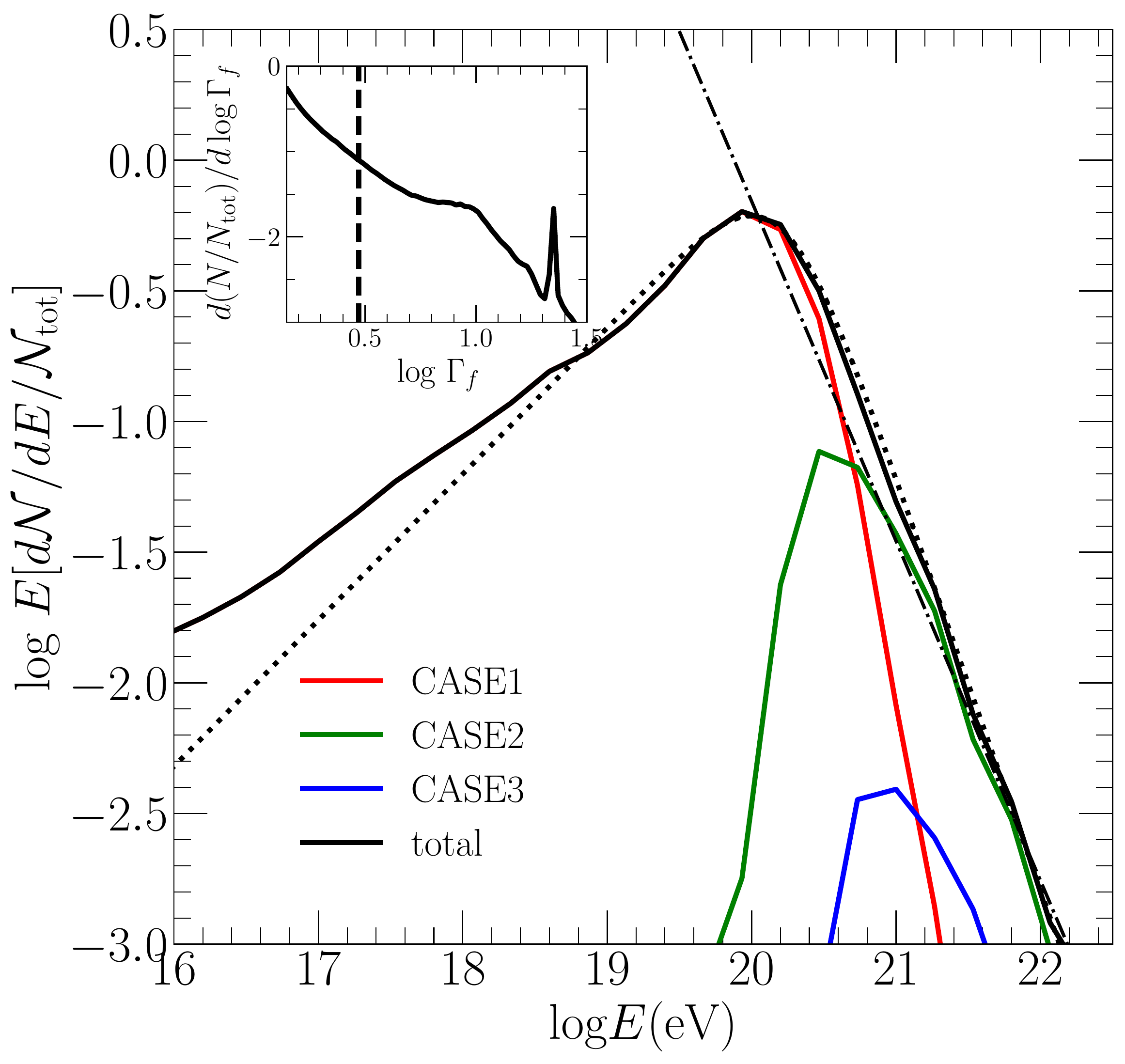}
\vskip -0.1 cm
\caption{Time-asymptotic energy spectra of particles belonging to CASE 1 (red), CASE 2 (green), and CASE 3 (blue), as well as all particles (black), for the high-resolution model Q46-$\eta5$-H. The black dotted line plots the double-power law fitting to the spectrum of all particles, and the black dot-dashed line draws the power law fitted to the high energy part of the CASE 2 spectrum. The spectrum of all particles and its fitting are identical to those in cyan in Figure \ref{f8}(b). The PDF of the Lorentz factor in the jet-spine flow is shown in the inserted box. The mean value, $\left<\Gamma_f\right>\approx3$, is indicated with the dashed line.}
\label{f10}
\end{figure}

Figure \ref{f9} shows the trajectories and their energization of three sample particles, one from each case, as a function of time since the injection. In the bottom panels, the red and blue lines follow the energy changes of the particles in the simulation and fluid frames, respectively. The two big jumps in the fluid frame energy (blue lines) for the CASE 2 (at $t-t_{\rm inj}\approx0.19$ and 0.27) and CASE 3 (at $t-t_{\rm inj}\approx0.21$ and 0.32) are a consequence of nGSA. Given that the jet-spine flow at the big-jump scattering points has $\Gamma_f\sim$ several, the energy gains in the jumps match the expectation due to scatterings into or out of the jet-spine flow. As the result of the energy boosts via nGSA, the particles of CASE 2 and CASE 3 reach well above $10^{20}$ eV. By contrast, the CASE 1 particle, which does not experience nGSA, fails to reach a very high energy.

In Figure \ref{f10}, we plot the time-asymptotic energy spectra for particles of CASE 1, CASE 2, and CASE 3, separately, in the case of the Q46-$\eta5$-H model. The total numbers of particle for each category are $\mathcal{N}_{\rm tot}(\rm CASE1)\gg \mathcal{N}_{\rm tot}(\rm CASE2)\gg \mathcal{N}_{\rm tot}(\rm CASE3)$. In CASE 1, particles are confined by the Hillas condition before they escape from the cocoon, so the spectrum (red) peaks at $E_{\rm max}$, above which it drops almost exponentially.

In CASE 2, particles that have already experienced nGSA scatter in the cocoon before they escape to the ICM. Note that the cocoon has an elongated shape (see Figure \ref{f1}), so particles even with $E\gtrsim E_{\rm max}$ can be confined lengthwise along the vertical direction, as shown in the top-middle panel of Figure \ref{f9}. The spectrum of the particles escaping from a cylinder with a finite radius and infinite length after isotropic scattering, is given as $d\mathcal{N}/dE\propto E^{-2}$. The spectrum of CASE 2 particles (green) has the {break} shifted to a higher energy due to nGSA, and follows a power-law distribution of slope $\sim-1.3$, or $d\mathcal{N}/dE\propto E^{-2.3}$ above $E_{\rm max}$. This is somewhat steeper than expected for the idealized setup, since the cocoon is not an infinitely stretched cylinder. The CASE 2 spectrum is the most dominant component in the high-energy part of the total spectrum. So the CASE 2 scenario would explain {the power-law spectrum above $E_{\rm break}$.}

CASE 3 particles are relatively rare, and hence their energy spectrum (blue) may not be accurately realized in our simulations. Nevertheless, the {break} of the spectrum shifts to a higher energy by about an order of magnitude, compared to that of CASE 1. Considering that the one cycle of nGSA produces a $\sim\Gamma_f^2$ boost in energy and the average Lorentz factor of the jet-spine flow is $\left<\Gamma_f\right> \approx 3$ (see the insert in Figure \ref{f10}), the shift of the {break} is well explained as a consequence of nGSA.

The combined spectrum of CASE 1, CASE 2, and CASE 3 results in the slope of $\sim-1.6$ above the {break}, which is a bit steeper than that of the CASE 2 spectrum, as demonstrated in Figure \ref{f10}.

{In Figure \ref{f8}(b), we compare the spectra of different resolution models for the Q46 jet. Compared to the fiducial case of Q46-$\eta5$-S (green), the spectrum of the higher resolution model, Q46-$\eta5$-H (cyan), shifts to higher energies, whereas the spectrum of the lower resolution model, Q46-$\eta5$-L (purple), shifts to lower energies. As mentioned in Section \ref{s2.3}, with a higher grid resolution, more significant nonlinear structures are induced; then, both DSA and TSA are more efficient owing to more frequent shocks and better developed turbulence in the cocoon. On the other hand, although $E_{\rm break}$ is slightly higher in higher resolution models, the resolution dependence seems to be not large.}

\subsection{Dependence on $B$, MFP, and Random Walk Models}\label{s5.3}

The {Monte Carlo (MC)} simulation results presented above are based on a number of modelings, such as those for the magnetic field strength in the jet-induced flows, the {mean-free-path (MFP)} of CR particles, and the random walk scheme. In this subsection, we briefly examine the effects of those modelings on the energy spectrum of escaping CRs. For that purpose, we use the Q46-$\eta5$-H jet.

\begin{figure}
\centering
\vskip 0.1 cm
\includegraphics[width=1\linewidth]{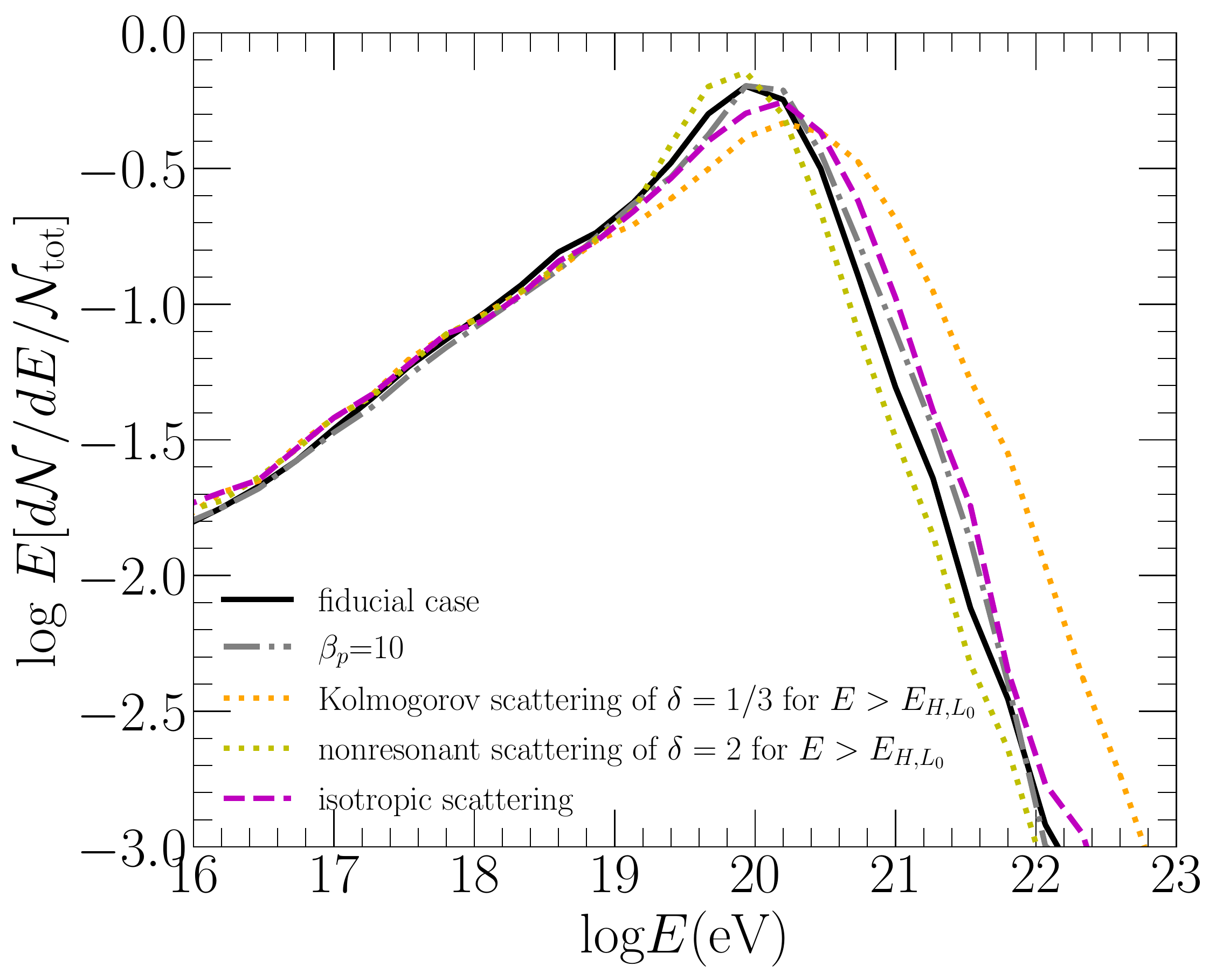}
\vskip -0.1 cm
\caption{Time-asymptotic energy spectra for the high-resolution model Q46-$\eta5$-H with different modelings for magnetic field and {Monte Carlo (MC)} simulations. The fiducial case (solid black) is compared to those of a stronger magnetic field with $\beta_p=10$ (dot-dashed gray), Kolmogorov scattering of $\delta=1/3$ for $E>E_{H,L_0}$ (dotted orange), nonresonant scattering of $\delta=2$ for $E>E_{H,L_0}$ (dotted dark yellow), and isotropic scattering (dashed magenta). The spectrum of the fiducial case is identical to the cyan curve in Figure \ref{f8}(b).}
\label{f11}
\end{figure}

We first examine the dependence on the magnetic field model described in Section \ref{s2.4}. In the relatively quiet flows, the magnetic field is likely to be prescribed by Equation (\ref{bie}); there, $B_p$ is stronger for lower $\beta_p$, which in turn may lead to more efficient particle acceleration. In contrast, in regions with well-developed turbulence, $B_{\rm turb}$ would dominate and the magnetic field is unaffected by the adopted value of $\beta_p$. In Figure \ref{f11}, the spectrum for the fiducial case ($\beta_p=100$, black solid line) is compared to that for the $\beta_p=10$ case (dot-dashed). The figure demonstrates that adopting a stronger magnetic field with smaller $\beta_p$ leads to slightly more efficient particle acceleration, but the difference is not large in the range of the magnetic field strength we consider.

As pointed in Section \ref{s3.1}, the nature of magnetic turbulence and also the diffusion/scattering of CR particles in the jet-induced flows are not completely understood, especially on scales larger than the coherence length of turbulence. Hence, we consider the three MFP models as listed in Table \ref{t2}. In Figure \ref{f11}, the energy spectra for the three models are displayed. Below $E_{H,L_0}\sim1$ EeV, with the same $\lambda_f$, the three spectra are basically identical. On the other hand, for $E>1$ EeV, the spectra differ for different MFP models. In Model A with $\delta=1/3$ (smaller $\lambda_f$), high-energy CRs go through more scatterings and tend to achieve higher energies, whereas in Model B with $\delta=2$ (larger $\lambda_f$), it works in the opposite way. Hence, the Model A spectrum shifts to higher energies and is also harder with a flatter slope above the {break}. In Model B, the spectrum shifts slightly to lower energies, while the slope above the {break} is nearly the same as in the fiducial model.

In Section \ref{s4.2}, we introduced a restricted random walk scenario through Equation (\ref{thetamax}). Here, we examine the dependence on this model by comparing the spectrum with isotropic scattering ($\psi\rightarrow\infty$) to the spectrum with restricted scattering ($\psi=1$, the fiducial case) in Figure \ref{f11}. As noted, the restricted, forward-beamed scattering reduces the frequency of the particle crossing across the shear interface, and hence suppresses the efficiency of nGSA. Hence, the spectrum of the fully isotropic scattering case shifts to higher energies, compared to that of the fiducial model, as expected.

\section{Summary}\label{s6}

We performed {relativistic hydrodynamic (RHD)} simulations for FR-II jets with the bulk Lorentz factor of $\Gamma_j\approx 7-70$, which propagate up to a few hundred kpc in the stratified ICM. Owing to the high-order, high-accuracy capabilities of our newly developed RHD code (Paper I), nonlinear structures such as shocks, turbulence, and shear are realized well enough to study diffusive shock acceleration (DSA), turbulent shear acceleration (TSA), gradual shear acceleration (GSA), and non-gradual shear acceleration (nGSA). As shown in Paper II, the overall jet morphology is governed mainly by the jet power $Q_j$. More powerful jets tend to generate more elongated cocoons, while less powerful jets develop broader cocoons full of mildly relativistic shocks and chaotic turbulence. The jet kinetic energy is dissipated mainly through shocks and turbulence in the jet-spine flow and the backflow. In addition, strong relativistic shear develops at the interface between the jet-spine flow and the backflow.

We then performed {Monte Carlo (MC)} simulations to study the transport and acceleration of CRs, utilizing the evolving snapshots of the jet-induced flow structures from the aforementioned RHD jet simulations. Toward this end, we adopted physically motivated recipes for the magnetic field in the jet flows, scattering MFP, and restricted random walks.

The main results are summarized as follows:

1. Injected CR particles are accelerated via the combination of DSA (shock), TSA (turbulence), and GSA/nGSA (shear), as they advect along the jet-spine flow and diffuse across the cocoon. CRs of $E\lesssim1$ EeV are energized mostly through DSA. Once they attain $E\gtrsim$ a few EeV, their MFP becomes comparable to the thickness of the shear layer between the jet-spine flow and the backflow, and GSA becomes important. Some of CRs with MFP large enough to cross the entire shear layer can be further accelerated via nGSA. Relativistic shear acceleration (including both GSA and nGSA) generates $\sim85\%$ of the total energy gain of {ultra-high-energy cosmic rays (UHECRs)} escaping from the system. DSA contributes only a few \% to the total energy gain, while TSA, although a subdominant process over the entire CR energy range, still generates about $10-15 \%$ (see Figure \ref{f6} and Table \ref{t3}).

2. The energy spectrum of all particles escaping from the jet approaches a time-asymptotic shape by the end of the simulations ($t_{\rm end}\sim 10^6-10^7$ yrs). The spectrum may be fitted to the double-power-law form given in Equation (\ref{spectfit}). The break energy, $E_{\rm break}$, can be interpreted by the size-limited maximum energy, $E_{\rm max}$, imposed by the width, $\mathcal{W}$, of the cocoon. $E_{\rm break}$ occurs at higher energies for higher power jets. On the other hand, the overall shape of the spectrum around the break shows only a weak dependence on the jet power. Just below $E_{\rm break}$, the spectrum follows $d\mathcal{N}/dE\propto E^{-0.5}$, which is expected when GSA and nGSA are the dominant processes. Above $E_{\rm break}$, it decreases as $d\mathcal{N}/dE\propto E^{-2.6}$, instead of the exponential cutoff. We interpret that this hard spectrum is a consequence of the nGSA boosts in CR energies at the shear interface between the jet-spine and the backflow and the anisotropic confinement in the elongated cocoon.

3. The time-asymptotic spectrum depends on various models and prescriptions employed in our MC simulations, such as those for $B$, $\lambda_f$, and $\delta \theta_{\rm max}$. However, the dependence seems to be marginal. Hence, although the spectrum presented in this paper may not be completely generalized, it should still provide a good measure and useful insights for the spectrum of UHECRs accelerated in FR-II radio jets.

In conclusion, we demonstrated that powerful radio galaxies equipped with the shear layer between the jet-spine flow and the backflow and the cocoon filled with shocks and turbulence could be potential cosmic accelerators of UHECRs well above $10^{20}$ eV. 

{Currently, we are carrying out similar RHD and MC simulations for FR-I radio galaxies, which are expected to make a significant contribution to the observed UHECR spectrum owing to their high number density \citep[e.g.][]{eichmann2022}. In particular, important local sources such as Centaurus A, Fornax A, and Virgo A are classified as FR-I radio galaxies. In a forthcoming study, we will explore if UHECRs arriving from both the local individual sources and the bulk populations of FR-I and FR-II radio galaxies could explain the observed energy spectrum and composition of UHECRs by considering their propagation through the intergalactic space.}

\begin{acknowledgments}
This work was supported by the National Research Foundation (NRF) of Korea through grants 2016R1A5A1013277, 2020R1A2C2102800, 2020R1F1A1048189, and RS-2022-00197685. Some of simulations were performed using the high performance computing resources of the UNIST Supercomputing Center.
\end{acknowledgments}

\bibliography{RadioGalaxy}{}
\bibliographystyle{aasjournal}

\end{document}